\documentclass[seceq]{ptptex}
\usepackage{graphicx} 
\usepackage{latexsym}
\notypesetlogo
\begin{document}
%\begin{flushright} 
%RUP- \\
%
%\end{flushright}
\vspace{10mm}
%\begin{document}
\begin{center}

\Large{ \bf Properties of solutions by the Schwinger-Dyson equation at finite 
 temperature and  density}

\vspace{5mm}

\Large{\bf $-$ A four-fermion interaction model $-$}

\vspace{15mm}
   
\large{ Hidekazu {\sc Tanaka} \footnote{E-mail:tanakah@rikkyo.ac.jp} and Shuji {\sc Sasagawa} \\
Rikkyo University, Tokyo 171-8501, Japan\\
}
 \end{center}

\begin{center}

\vspace{25mm}

{\Large ABSTRACT}
 \end{center}
        
\vspace{10mm}

In this paper, we examine the properties of the solutions obtained by the Schwinger-Dyson equation (SDE).
 As a simple example, we consider a two-dimensional model including four-fermion interaction.  It is shown that when this model is solved by an iterative method using the SDE at  finite density, multiple solutions depending on the initial input values  are obtained.
 We investigate the reasons for this situation by examining the convergence of the solutions obtained by iterative method. We also consider the solution of the SDE using alternative methods. Furthermore, we compare these results with analytical solutions at zero temperature and discuss the relation with the behavior of the effective potential.  We extend these considerations to finite temperatures. 
 Based on these analyses, we consider ways to avoid spurious solutions that appear in the iterative method.

%\markboth{H. {\sc Tanaka}   }{}
\def\proj{{\bf P}}
\def\slsh#1{{#1}{\kern-6pt}/{\kern1pt}}
%\title{ }

%\vspace{15mm}

%\author{Hidekazu {\sc Tanaka}}
%\inst{ Rikkyo University, \\
%           Nishi-ikebukuro, Toshima-ku Tokyo, Japan, 171 \\
%     }
%\recdate{ }
%  \begin{center}
%\maketitle
%\vspace{25mm}

%  {\Large ABSTRACT}
%  \end{center}
      
%\vspace{10mm}

% \maketitle

%  \newpage

\newpage

\section{Introduction}

The physics of strongly coupled quantum systems is a very interesting research subject.  However, it is generally difficult to solve the physical systems in the strongly coupled region, such as quantum chromodynamics (QCD) in the infrared region, in which the approximate chiral symmetry is broken due to non-perturbative effects and the light quarks have large effective masses.(See reviews in Refs [1-3].)  

The most powerful method to numerically solve non-perturbative systems is simulation in lattice theory. But unfortunately, at present, the numerical simulations using the lattice theory have difficulty in evaluating the region for low-temperature at high-density states due to the sign problem. (See recent review [4]. Related references can be found there.)
 
The Schwinger-Dyson equation (SDE) [5,6] is a method used to evaluate non-perturbative phenomena.
 The SDE is a nonlinear integral equation derived from quantum field theory.
  If the SDE cannot be solved analytically, a method of iteratively solving the SDE starting from an arbitrarily given input value is often used [2].
 The iterative method is not limited to solving the SDE, but is generally used when solving nonlinear equations numerically.

In this paper, we investigate the behavior of the solutions when solving the SDE for finite temperature and  density systems using the iterative method.
 As an example, we examine the contact four-fermion interaction model in $(1+1)$ space-time dimensions, such as the Gross-Neveu (GN) model for chiral fermions at large flavor limit [7].  In this model, chiral symmetry is broken and massless fermions have an effective mass.   
  So far, there have been many studies on the GN model. For example, the GN model was extended to finite temperature [8-11] and density [12-14].
  This type of model is often used as a simplified model of QCD.
 
We show that when we iteratively solve the SDE of our model in the case of finite density, multiple solutions that depend on the initial input values appear. 
This suggests that the numerical results contain spurious solutions that depend on the initial input values. In this paper, we will discuss how to understand this phenomenon. Since our model is relatively easy to solve analytically in the low temperature limit, we can investigate the numerical solutions by comparing them with the analytical solutions.
 
This paper is organized as follows:
In Sect.2, we formulate the SDE for the effective mass of fermions at finite temperature and density in our model.  In Sect.3, we use the iterative method to numerically calculate the effective mass with several initial input values, 
and then investigate the properties of the numerical solutions by comparing them with the analytical solutions at zero temperature and finite density. 
Furthermore, we solve the SDE using another method, and discuss the relation between the SDE solutions and the effective potential. In Sect.4, we investigate the solutions of the SDE using the iterative method at finite temperature and density. In Sect.5, based on our analyses, we consider how spurious solutions can be avoided.  Sect.6 is devoted to a summary and some comments.
Appendices A and B provide further details on the properties of solutions that satisfy the SDE and on the convergence properties of the iterative method for the SDE.

\section{The SDE for a four-fermion interaction model}

In this section, we formulate the SDE for the effective mass $M$ at temperature $T$ and chemical potential $\mu$ in the model with four-fermion interaction for chiral fermions  in  $(1+1)$ space-time dimensions, such as the GN model. \footnote{The Lagrangian density of the GN model is given by $$ {\mathcal L}=i\sum_{k=1}^{N}{\bar \psi}^{(k)}\slsh{\partial}\psi^{(k)}+{g^2 \over 2}\left(\sum_{k=1}^{N}{\bar \psi}^{(k)}\psi^{(k)}\right)^2 $$
using $N$ two-component massless spinors in $(1+1)$ space-time dimensions $\psi^{(k)}$ with $k=1\cdots N$. The second term is the interaction term between four fermions. 
Equation (2$\cdot$1) is the gap equation obtained in the large $N$ limit[7].
Here, the coupling constant is defined as $\lambda =2g^2N/(2\pi)^2$.}
 We start from the SDE for the model at $T=\mu=0$ as
\begin{eqnarray}
M=i\lambda \int d^2Q{M \over Q^2-M^2+i\varepsilon}
=i\lambda \int dq_0dq{M \over q_0^2-q^2-M^2+i\varepsilon}
\end{eqnarray}
with a momentum $Q=(q_0,q)$ and $\lambda$ is a coupling constant.
 Here, using the standard method [15], we extend Eq.(2$\cdot$1) to finite temperature and density in the imaginary-time formalism, in which we replace $q_0$ to $q_0=i\omega_n+\mu$ with  the Matsubara frequency $\omega_n=2\pi T(n+1/2) ~(n=0,\pm 1,\pm 2, \cdots$) for fermions and integration over  $q_0$ to $2\pi T i\sum_{n=-\infty}^{\infty} $, respectively.
Then, Eq.(2$\cdot$1) is given as 
$$ M(T,\mu)=i\lambda (2\pi T i)\sum_{n=-\infty}^{\infty}\int dq{M(T,\mu) \over (i\omega_n+\mu)^2-q^2-M^2(T,\mu)+i\varepsilon}$$
 \begin{eqnarray}
\equiv \lambda \int dq M(T,\mu)I(T,\mu,q)
\end{eqnarray}
with
\begin{eqnarray}
I(T,\mu,q)= 2\pi T \sum_{n=-\infty}^{\infty} {1 \over  -(i\omega_n+\mu)^2+q^2+M^2(T,\mu)-i\varepsilon}.
\end{eqnarray}
Defineing $E^2(T,\mu,q)=q^2+M^2(T,\mu)-i\varepsilon$, we write
$$I(T,\mu,q)
\equiv -2\pi T \sum_{n=-\infty}^{\infty} {1 \over  q_0^2-E^2(T,\mu,q)}$$
\begin{eqnarray}
=-2\pi T \sum_{n=-\infty}^{\infty} {1 \over  (q_0-E(T,\mu,q))(q_0+E(T,\mu,q))}\equiv T \sum_{n=-\infty}^{\infty} g(q_0 = i\omega_n+\mu)
\end{eqnarray}
with
\begin{eqnarray}
g(q_0 = i\omega_n+\mu) = -2\pi  {1 \over  (q_0-E(T,\mu,q))(q_0+E(T,\mu,q))}.
\end{eqnarray}
In the following calculations, we define the energy as $E(T,\mu,q)=\sqrt{q^2+M^2(T,\mu)}$ and ignore the $i\varepsilon$ term.

 Rewriting the last expression of Eq.(2$\cdot$4) using an integral form with the integral path $C$ surrounding all poles  of the function $\tanh(\beta(q_0-\mu)/2)$ in the complex $q_0$ plane as 
\begin{eqnarray}
T\sum_{n=-\infty}^{\infty}g(q_0=i\omega_n+\mu)=\oint_{C} {dq_0 \over 2\pi i}g(q_0){1\over 2}\tanh\left({1 \over 2}\beta (q_0-\mu)\right),
\end{eqnarray}
we have 
$$
 I(T,\mu,q)=-i\oint_{C} {dq_0 \over 2\pi}g(q_0){1 \over 2}\tanh\left({\beta(q_0-\mu) \over 2}\right) $$
\begin{eqnarray}
=i\oint_{C} dq_0 {1 \over  (q_0-E(T,\mu,q))(q_0+E(T,\mu,q))}{1 \over 2}\tanh\left({\beta(q_0-\mu) \over 2}\right)
\end{eqnarray}
with $\beta=1/T$. The integral path $C$ can be converted into two integral paths $C_1$ and $C_2$, where $C_1$ is the integral path from $\mu+\eta-i\infty$ to $\mu+\eta+i\infty$, and $C_2$ is the integral path from $\mu-\eta+i\infty$ to $\mu-\eta-i\infty$ , respectively,  with $\eta\rightarrow 0$.
 Furthermore, by closing the integral paths $C_1$ and $C_2$ with large half-circles in the complex $q_0$ plane, we can obtain the residues in the integral given by Eq.(2$\cdot$7).

From these calculations, the SDE can be written as
$$ M(T,\mu)=\lambda\int dq M(T,\mu)I(T,\mu,q)$$
\begin{eqnarray}
=\pi\lambda\int dq{ M(T,\mu) \over  2E(T,\mu,q)}\left[\tanh\left({\beta E^{(-)}(T,\mu,q)\over 2}\right)
+\tanh\left({\beta E^{(+)}(T,\mu,q)\over 2}\right)\right].
\end{eqnarray}
 Here, we define $E^{(\pm)}(T,\mu,q)=E(T,\mu,q)\pm \mu$.
 
 Introducing a momentum cutoff $\Lambda$, we write  the SDE as
$$M(T,\mu)= \pi \lambda\int_{-\Lambda}^{\Lambda}  dq { M(T,\mu) \over  E(T,\mu,q)} \zeta(T,\mu,E(T,\mu,q)) $$
\begin{eqnarray}
=2\pi \lambda\int_{0}^{\Lambda}  dq { M(T,\mu) \over  E(T,\mu,q)} \zeta(T,\mu,E(T,\mu,q))
\end{eqnarray}
with 
\begin{eqnarray}
\zeta(T,\mu,E(T,\mu,q))\equiv {1 \over 2}\left[\tanh\left({\beta E^{(-)}(T,\mu,q)\over 2}\right)+\tanh\left({\beta E^{(+)}(T,\mu,q)\over 2}\right)\right].
\end{eqnarray}
 
In the following calculations, $M(T,\mu),\mu$ and $E(T,\mu,q)$ are assumed to be positive values.

\section{Analytical and numerical solutions at $T=0$}

In this section, we solve the SDE given by Eq.(2$\cdot$9) with Eq.(2$\cdot$10) at $T=0$. We first show the results of calculating the effective mass using the iterative method.
Next, we find the effective mass analytically.
In order to understand the multiple solutions obtained by the iterative method, we examine the convergence properties of the solutions near the analytical solutions. Furthermore, we will show another method of finding the solutions of the SDE, and finally discuss the relation between the effective potential and the solutions obtained by the SDE.

Here, in the limit $T\rightarrow 0 (\beta\rightarrow \infty)$, we use $\zeta(T,\mu,E(T,\mu,q))\rightarrow \Theta(E(0,\mu,q)-\mu)$.\footnote{$\Theta(x)$ denotes the Heaviside step function.}\footnote{$M(0,\mu)$ and $E(0,\mu,q)$ denote the effective mass $M(T=0,\mu)$ and the energy $E(T=0,\mu,q)$, respectively. } 
 Therefore, the SDE at $T=0$ is given by 
\begin{eqnarray}
 M(0,\mu)=2\pi\lambda\int_{0}^{\Lambda} dq{ M(0,\mu) \over  E(0,\mu,q)}\Theta(E(0,\mu,q)-\mu).
 \end{eqnarray}

\subsection{Solutions of the SDE obtained by the iterative method }

In this subsection, we solve the SDE given by Eq.(3$\cdot$1) using the iterative method at $T=0$.\footnote{In actual numerical calculations, the momentum integral is performed using $\Lambda\geq q\geq \delta$ with $\delta/\Lambda=5\times 10^{-3}$ to avoid numerical instabilities at $M(0,\mu)\rightarrow 0$.  This restriction has almost no effect on the numerical results in the region of $M(0,\mu)\neq 0$.} Here, the coupling constant is taken to be $\lambda=0.3$.
 
Fig.1 shows the solutions for several different initial input values $M_0$. 

\begin{figure}
\centerline{\includegraphics[width=8cm]{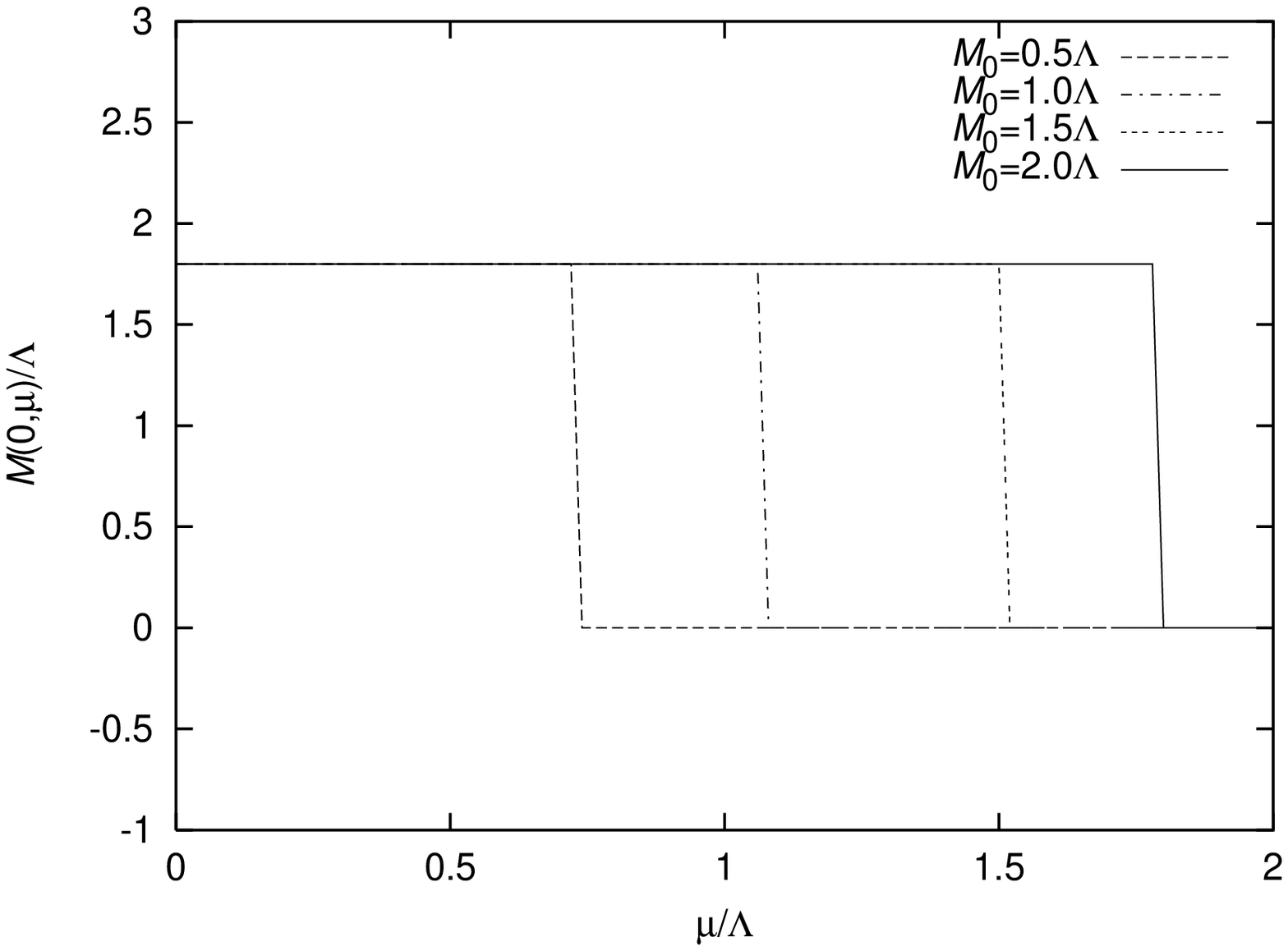}}
\caption{The chemical potential dependences  of the effective masses at $T=0$
 for several different initial input values $M_0$  when solving the SDE using the iterative method. Here, the coupling constant is taken to be $\lambda=0.3$. The horizontal axis shows the chemical potential divided by $\Lambda$.}
%\label{Fig.1}
\end{figure}

From Fig.1, it can be seen that the SDE solutions obtained by the iterative method depend on the initial input values. Here, $M(0,\mu)\rightarrow 0$ is usually interpreted as the restoration of the chiral symmetry of the fermions, but the phase transition point seems to depend on the initial input values.

\subsection{Analytical solutions of the SDE }

In this subsection, we analytically solve the SDE given in Eq.(3$\cdot$1).

 For $M(0,\mu)\neq 0$, we will find the effective mass $M(0,\mu)$ that satisfies\begin{eqnarray} 
1=2\pi\lambda\int_{0}^{\Lambda} dq{ 1 \over  E(0,\mu,q)}\Theta(E(0,\mu,q)-\mu)
\end{eqnarray}
with $E(0,\mu,q)=\sqrt{q^2+M^2(0,\mu)}$.

Due to the restriction of $E(0,\mu,q)>\mu$, the range of the $q^2$   becomes $\Lambda^2 \geq q^2 > \mu^2-M^2(0,\mu).$
 From this relation, when $\mu < M(0,\mu)$, the integration range of $q$  in the SDE  is $0 \leq q \leq \Lambda$. On the other hand, when $\mu > M(0,\mu)$, the integration range of $q$ in the SDE is given by $q_{\rm L}\equiv \sqrt{\mu^2-M^2(0,\mu)} < q \leq \Lambda$.

For $\mu < M(0,\mu)$, Eq.(3$\cdot$2) becomes
$$ 1=2\pi\lambda\int_0^{\Lambda} dq{ 1 \over  E(0,\mu,q)}
= 2\pi\lambda\int_0^{\Lambda} dq{ 1 \over  \sqrt{q^2+M^2(0,\mu)}}$$
\begin{eqnarray}=2\pi\lambda\log{\Lambda+\sqrt{\Lambda^2+M^2(0,\mu)} \over M(0,\mu)}.\end{eqnarray}
 With $m(0,\mu)=M(0,\mu)/\Lambda$ and 
\begin{eqnarray}\xi\equiv e^{1/(2\pi\lambda)}={\Lambda+\sqrt{\Lambda^2+M^2(0,\mu)} \over M(0,\mu)}
={1+\sqrt{1+m^2(0,\mu)} \over m(0,\mu)}\end{eqnarray}
from Eq.(3$\cdot$3),  Eq.(3$\cdot$4) is given as
\begin{eqnarray}m(0,\mu)\xi-1=\sqrt{1+m^2(0,\mu)}.
\end{eqnarray}
When both sides of the above equation are squared,  we have  
\begin{eqnarray}m(0,\mu)[ m(0,\mu)\xi^2-2\xi-m(0,\mu)]=0.\end{eqnarray}
For $ m(0,\mu)\neq 0$, the solution is given as 
\begin{eqnarray} m^{(1)}(0,\mu)=M^{(1)}(0,\mu)/\Lambda={2\xi \over \xi^2-1},\end{eqnarray}
which does not depend on $\mu$. 

On the other hand, for $\mu > M(0,\mu)$, Eq.(3$\cdot$2) becomes
$$  1=2\pi\lambda\int_{q_{\rm L}}^{\Lambda} dq{ 1 \over  E(0,\mu,q)}
= 2\pi\lambda\int_{q_{\rm L}}^{\Lambda} dq{ 1 \over  \sqrt{q^2+M^2(0,\mu)}}
$$
\begin{eqnarray} 
=2\pi\lambda\log{\Lambda+\sqrt{\Lambda^2+M^2(0,\mu)} \over \sqrt{\mu^2-M^2(0,\mu)}+\mu}.
\end{eqnarray}
With $m(0,\mu)=M(0,\mu)/\Lambda$, ${\bar \mu}=\mu/\Lambda$, ${\bar \Delta}\equiv q_{\rm L}/\Lambda=\sqrt{{\bar \mu}^2-m^2(0,\mu)}$ and 
\begin{eqnarray}
 \xi= e^{1/(2\pi\lambda)}={\Lambda+\sqrt{\Lambda^2+M^2(0,\mu)}\over \sqrt{\mu^2-M^2(0,\mu)}+\mu}
={1+\sqrt{1+m^2(0,\mu)} \over \sqrt{{\bar \mu}^2-m^2(0,\mu)}+{\bar \mu}}
\end{eqnarray}
from Eq.(3$\cdot$8), Eq.(3$\cdot$9) is given as 
\begin{eqnarray}
\xi\left[{\bar \Delta}+{\bar \mu}\right]-1 = \sqrt{1+{\bar \mu}^2-{\bar \Delta}^2}.
\end{eqnarray}
By squaring both sides of Eq.(3$\cdot$10), we have a quadratic equation for ${\bar \Delta}$ as
\begin{eqnarray}
(\xi^2+1){\bar \Delta}^2 +2{\bar \Delta}(\xi^2{\bar \mu}-\xi)
+\xi^2{\bar \mu}^2-2\xi{\bar \mu}-{\bar \mu}^2=0.
 \end{eqnarray}
The solutions to this equation are given by 
\begin{eqnarray}
{\bar \Delta}_{\pm}={-\xi(\xi{\bar \mu}-1) \pm (\xi+{\bar \mu}) \over \xi^2+1}, \end{eqnarray}
which give
\begin{eqnarray}
{\bar \Delta}_{+}={-\xi(\xi{\bar \mu}-1) + (\xi+{\bar \mu}) \over \xi^2+1} 
={2\xi -(\xi^2-1){\bar \mu} \over \xi^2+1} 
\end{eqnarray}
and
\begin{eqnarray}
{\bar \Delta}_{-}={-\xi(\xi{\bar \mu}-1) - (\xi+{\bar \mu}) \over \xi^2+1} 
={ -(\xi^2+1){\bar \mu} \over \xi^2+1}=-{\bar \mu} ,
\end{eqnarray}
respectively. Since ${\bar \mu}\geq 0$, ${\bar \Delta}_-$ is not a solution that satisfies ${\bar \Delta}> 0$. Also, from ${\bar \Delta}_+> 0$, the possible range of ${\bar \mu}$ is given by
\begin{eqnarray}
{\bar \mu} < {2\xi \over \xi^2-1}=m^{(1)}(0,\mu). 
\end{eqnarray}
In the above range of ${\bar \mu}$, the second solution of the SDE  becomes 
\begin{eqnarray}
 m^{(2)}(0,\mu)=M^{(2)}(0,\mu)/\Lambda=\sqrt{{\bar \mu}^2-{\bar \Delta}_{+}^2}
={2\sqrt{\xi(\xi+{\bar \mu})(\xi{\bar \mu}-1)}\over (\xi^2+1)}.
 \end{eqnarray}
 For $M(0,\mu)=\mu$, $q_{\rm L}=0$ holds. Therefore Eqs.(3$\cdot$3) and (3$\cdot$8) have the same solution.
Also, since $m(0,\mu)={\bar \mu}$, we have the relation $m^{(1)}(0,\mu)=m^{(2)}(0,\mu)={\bar \mu}$.

The analytical solutions obtained at $T=0$ and $\lambda=0.3$ are plotted in Fig.2. The solution $m^{(1)}(0,\mu)$ for $m(0,\mu)>{\bar \mu}$ shown in Eq.(3$\cdot$7) is denoted by the solid line, and the solution $m^{(2)}(0,\mu)$ for $m(0,\mu)<{\bar \mu}$ shown in Eq. (3$\cdot$16) is denoted by the dashed-dotted curve.

\begin{figure}
\centerline{\includegraphics[width=8cm]{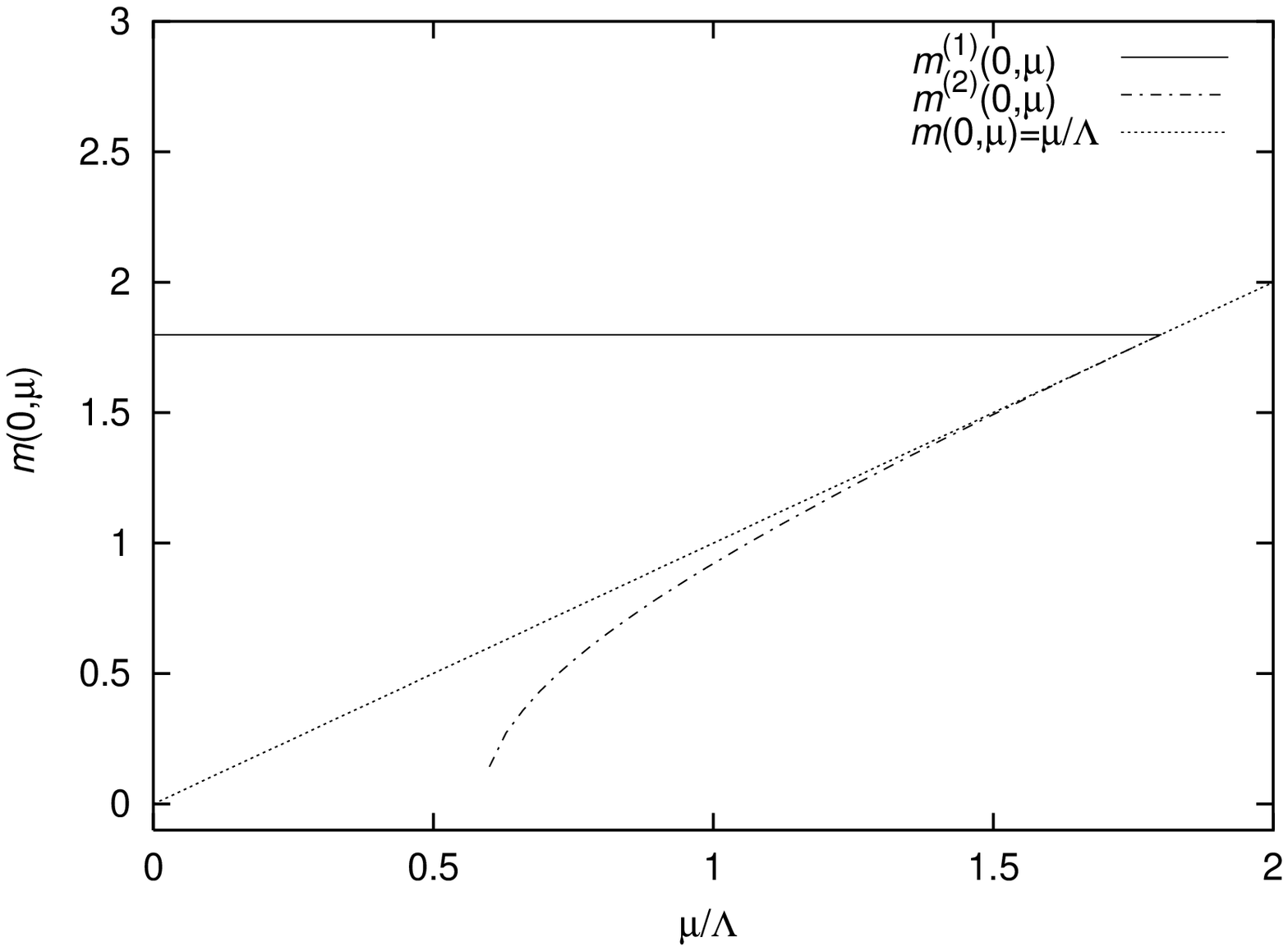}}
\caption{The chemical potential dependences of the analytical solutions that satisfy Eq.(3$\cdot$2) at $T=0$. The solid line and the dashed-dotted curve denote the analytical solutions $m^{(1)}(0,\mu)$ and $m^{(2)}(0,\mu)$, respectively, at $\lambda=0.3$. The horizontal axis shows the chemical potential divided by $\Lambda$.}
%\label{Fig.2}
\end{figure}

Fig.2 shows that the two solutions come together at $m(0,\mu)={\bar \mu}=2\xi/(\xi^2-1)\simeq 1.80$, and there is no solution in larger region of ${\bar \mu}$. On the other hand, from Eq.(3$\cdot$16), we have $m^{(2)}(0,\mu)\rightarrow 0$ for ${\bar \mu}\rightarrow 1/\xi \simeq 0.59$. Here, $m(0,\mu)=0$ is the region where chiral symmetry is restored. Since the solutions we have found correspond to the solutions for $M(0,\mu)\neq 0$, the point $m^{(2)}(0,\mu)=0$ is outside the scope of our current discussion.

\subsection{Convergence properties of the solutions obtained by the iterative method}

In order to understand the behavior shown in Fig.1, Fig.3 and Fig.4 show the convergence properties of the solutions when solving the SDE by the iterative method starting from initial input values near the analytical solutions of the effective mass at ${\bar \mu}=0.80$ and $\lambda=0.3$.

\begin{figure}
\centerline{\includegraphics[width=8cm]{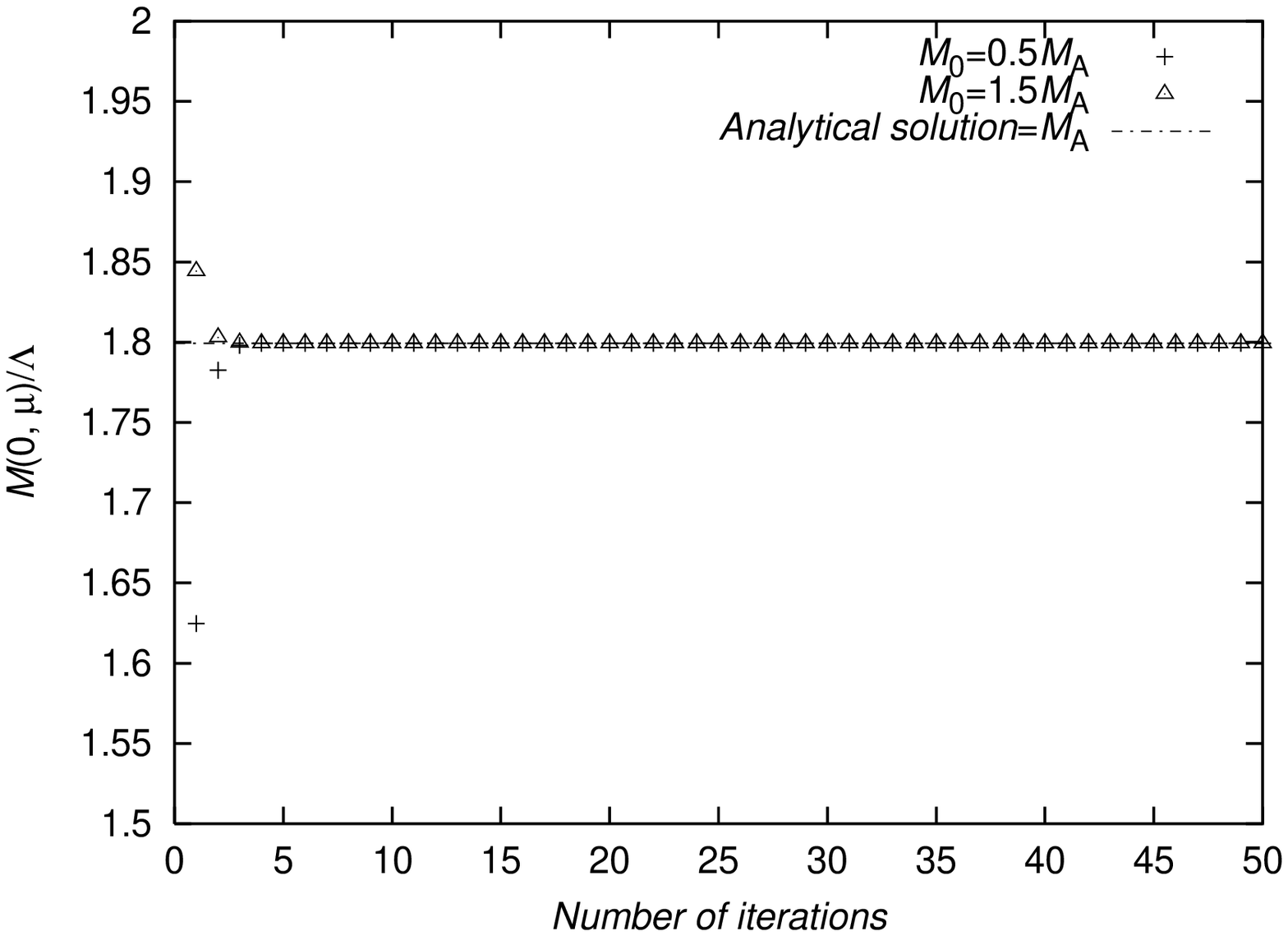}}
\caption{The initial input value dependences of convergence of the effective mass at ${\bar \mu}=0.80$ and $\lambda=0.3$. The dash-dotted line indicates the value of the analytical solution $M_{\rm A}/\Lambda=m^{(1)}(0,\mu)\simeq 1.80$ at ${\bar \mu}=0.80$. The cross symbols indicate the case where the initial input value is 0.5 times $m^{(1)}(0,\mu=0.80\Lambda)$, and the triangle symbols indicate the case where the initial input value is 1.5 times $m^{(1)}(0,\mu=0.80\Lambda)$.
The horizontal axis shows the number of iterations.}
%\label{Fig.3}
\end{figure}

\begin{figure}
\centerline{\includegraphics[width=8cm]{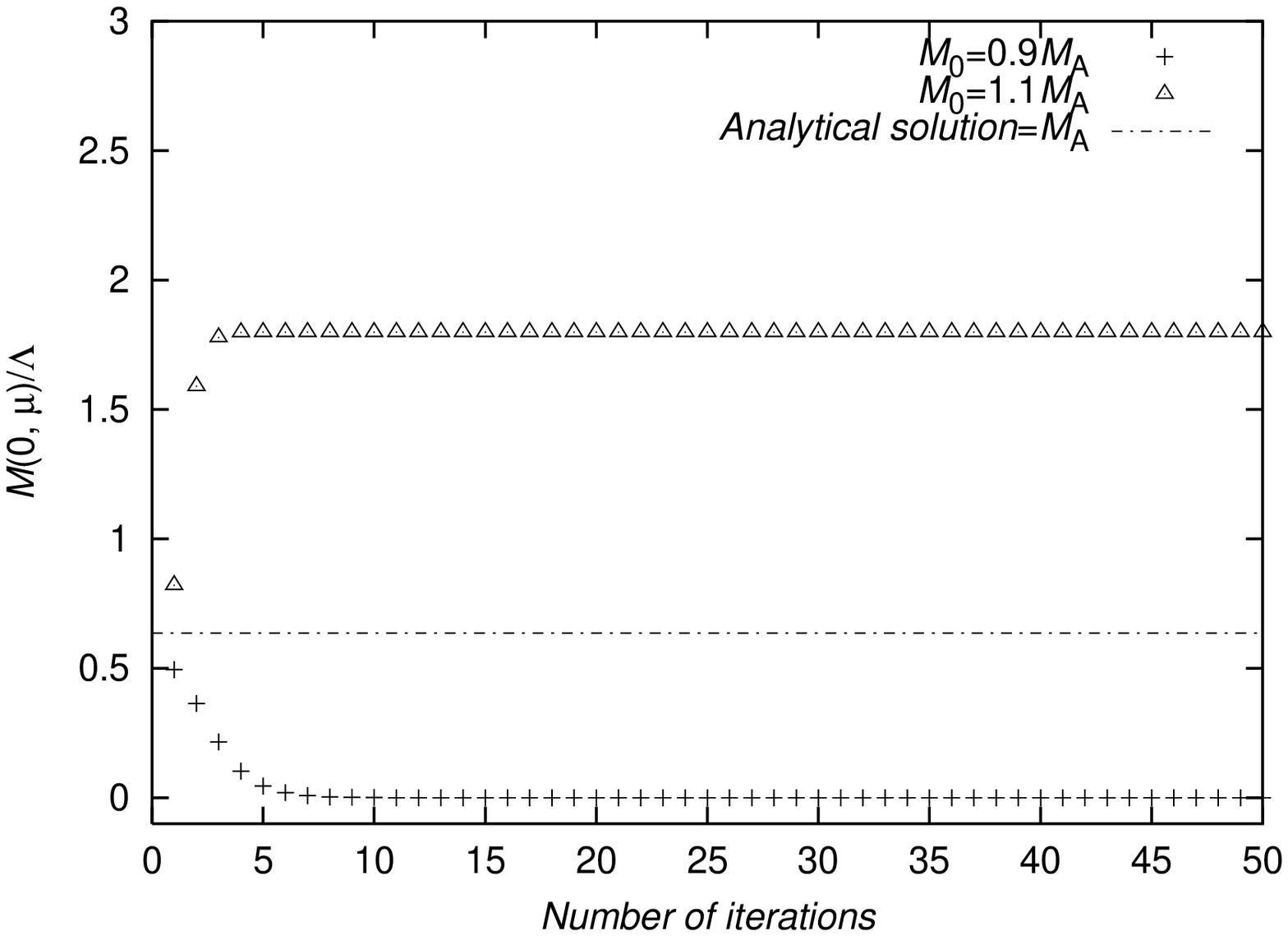}}
\caption{The initial input value dependences of convergence of the effective mass at ${\bar \mu}=0.80$ and $\lambda=0.3$. The dash-dotted line indicates the value of the analytical solution $M_{\rm A}/\Lambda=m^{(2)}(0,\mu)\simeq 0.64$ at ${\bar \mu}=0.80$. The cross symbols indicate the case where the initial input value is 0.9 times $m^{(2)}(0,\mu=0.80\Lambda)$, and the triangle symbols indicate the case where the initial input value is 1.1 times $m^{(2)}(0,\mu=0.80\Lambda)$.
The horizontal axis shows the number of iterations.}
%\label{Fig.4}
\end{figure}

 From Fig.3, when the initial input values are larger than $\mu$, the effective mass converges to $m^{(1)}(0 ,\mu)$. In Fig.3, $M_{\rm A}/\Lambda=m^{(1)}(0,\mu)\simeq 1.80$, so the initial input values $M_0$ are $1.5M_{\rm A}\simeq 2.7\Lambda$ and $0.5M_{\rm A}\simeq 0.90\Lambda$, which are larger than $\mu=0.80\Lambda$.
 From Fig.4, when the initial input value is larger than $m^{(2)}(0,\mu)$, the effective mass converges to $m^{(1)}(0 ,\mu)$.
 On the other hand, if the initial input value is smaller than $m^{(2)}(0,\mu)$, the effective mass converges to zero. In Fig.4, $M_{\rm A}/\Lambda=m^{(2)}(0,\mu)\simeq 0.64$, so the initial input values $M_0$ are $1.1M_{\rm A}\simeq 0.70\Lambda$ and $0.9M_{\rm A}\simeq 0.58\Lambda$, which are smaller than $\mu=0.80\Lambda$.
 
 The following can be seen from the convergences of the solutions obtained by  the SDE using the iterative method. 
When $m^{(2)}(0,\mu)$ increases as $\mu$ increases and $m^{(2)}(0,\mu)$ exceeds the initial input value, the effective mass converges to zero.
Therefore, it should be noted that the behavior shown in Fig.1 does not represent the restoration of the chiral symmetry except for the solution obtained with 
the initial input mass $M_0 /\Lambda =2.0$.

This behavior may be understood as follows. 
For $\mu>M(0,\mu)$, the SDE can be written as 
\begin{eqnarray} 
M(0,\mu)=2\pi\lambda M(0,\mu)\log{\Lambda+\sqrt{\Lambda^2+M^2(0,\mu)} \over \sqrt{\mu^2-M^2(0,\mu)}+\mu}
\end{eqnarray}
from Eq.(3$\cdot$8).
Substituting the initial value $M(0,\mu)=M_0$ into the right side of the above equation to find the next solution $M_1$, the condition for $M_1<M_0$ is 
\begin{eqnarray} 
1 > {M_1 \over M_0}=2\pi\lambda \log{\Lambda+\sqrt{\Lambda^2+M^2_0} \over \sqrt{\mu^2-M^2_0}+\mu}.
\end{eqnarray}
Solving Eq.(3$\cdot$18) in a similar  way as we used to obtain (3$\cdot$16) yields $M_0 < M^{(2)}(0,\mu)$.
Therefore, when the input value $M_0$ is smaller than $M^{(2)}(0,\mu)$, the solution obtained by repeatedly using Eq.(3$\cdot$17) decreases and converges to $M(0,\mu)\rightarrow 0$.
On the other hand, when the input value $M_0$ is larger than $M^{(2)}(0,\mu)$, the solution obtained by repeatedly using Eq.(3$\cdot$17) increases and $M(0,\mu)$ becomes larger than $\mu$.

For $M(0,\mu)>\mu$, the SDE is given by
\begin{eqnarray} 
M(0,\mu)=2\pi\lambda M(0,\mu)\log{\Lambda+\sqrt{\Lambda^2+M^2(0,\mu)} \over M(0,\mu)}
\end{eqnarray}
from Eq.(3$\cdot$3). Substituting the initial value $M(0,\mu)=M_0$ into the right side of the above equation to find the next solution $M_1$, the condition for $M_1<M_0$ is given as
\begin{eqnarray} 
1 > {M_1 \over M_0}= 2\pi\lambda \log{\Lambda+\sqrt{\Lambda^2+M^2_0} \over M_0}.
\end{eqnarray}

Solving Eq.(3$\cdot$20) in a similar  way as we used to obtain (3$\cdot$7) yields $M_0 > M^{(1)}(0,\mu)$.
 Therefore, when the input value $M_0$ is larger than $M^{(1)}(0,\mu)$, the solution obtained by repeatedly using Eq.(3$\cdot$19) decreases.
 On the other hand, when the input value $M_0$ is smaller than $M^{(1)}(0,\mu)$, the solution obtained by repeatedly using Eq.(3$\cdot$19) increases.
 Therefore, by repeating the calculation using the SDE in the region of $M(0,\mu)>\mu$, we finally obtain the solution that converges to $M^{(1)}(0,\mu)$.

From the above considerations, by repeating calculations using the SDE, when the initial input value $M_0$ is $M_0<M^{(2)}(0,\mu)$, the effective mass converges to zero. On the other hand, when $M_0$ is $M_0>M^{(2)}(0,\mu)$, the effective mass converges to $M^{(1)}(0,\mu)$.

\subsection{Another method for solving the SDE}

We will try to solve the SDE using another method in which 
we  numerically find the relation between $M(0,\mu)$ and $\mu$ that satisfies the equality in Eq.(3$\cdot$2). Here, we define
\begin{eqnarray}
1=2\pi\lambda\int_{0}^{\Lambda} dq{ 1 \over  E(0,\mu,q)}\Theta(E(0,\mu,q)-\mu)\equiv F(M,\mu,T=0)
\end{eqnarray}
with $E(0,\mu,q)=\sqrt{q^2+M^2(0,\mu)}$ and $M$ in the function $F(M, \mu, T=0)$ indicates $M(0,\mu)$.

Fig.5 shows the effective mass dependences of $F(M,\mu,T=0)$ defined in Eq.(3$\cdot$21) for several different values of chemical potential at $\lambda=0.3$. 
 When changing $M$, $M$ which satisfy $F(M,\mu,T=0)=1$ are the solutions of the SDE at $T=0$.
\begin{figure}
\centerline{\includegraphics[width=8cm]{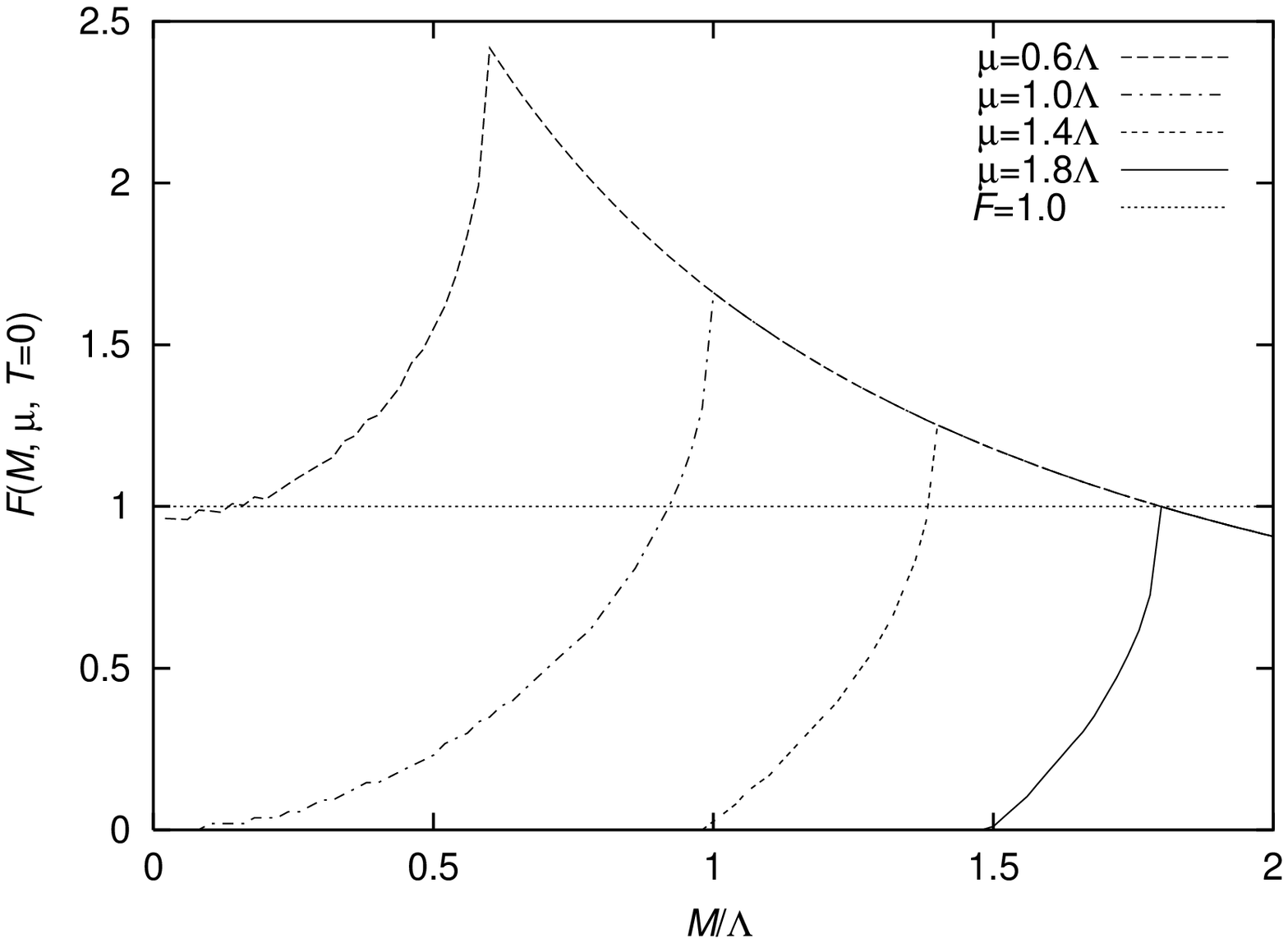}}
\caption{The effective mass dependences of $F(M,\mu,T=0)$  at $\lambda=0.3$ for  several different values of chemical potential. $M$ which satisfy $F(M,\mu,T=0)=1$ are the solutions of the SDE at $T=0$.
 The horizontal axis shows the effective mass divided by $\Lambda$.}
%\label{Fig.5}
\end{figure}

From Fig.5, we can see that there are two solutions for each $\mu$ when $\mu < 1.80\Lambda$. Here, the larger and smaller values of $M/\Lambda$ that satisfy $F(M,\mu,T=0)=1$ correspond to the solutions $m^{(1)}(0, \mu)=M^{(1)}(0,\mu)/\Lambda$ and $m^{(2)}(0,\mu)=M^{(2)}(0,\mu)/\Lambda$, respectively.
For $\mu\simeq 1.80\Lambda$, there is only one solution that satisfies $F(M,\mu,T=0)=1$, which corresponds to the point $m^{(1)}(0,\mu)=m^{(2)}(0,\mu)\simeq 1.80$ in Fig. 2. 
On the other hand, when $\mu> 1.80\Lambda$, the solution that satisfies $F(M,\mu,T=0)=1$ does not exist. 

\subsection{Correspondence between the SDE solutions and the effective potential }

To further understand the meaning of the solutions obtained by the SDE, we examine the effective potential corresponding to the SDE. Using $V(M,\mu,T)$ \footnote{$M$ indicates $M(T,\mu)$.} as the effective potential, the SDE is derived from
\begin{eqnarray} {\partial V(M,\mu,T) \over \partial M(T,\mu)}=0.\end{eqnarray}
 The effective potential that reproduces Eq.(3$\cdot$1) can be written as [12-14] 
$$V(M,\mu,T)=-4\int_{0}^{\Lambda} dq[E(T,\mu,q)+T\log(1+e^{-\beta(E(T,\mu,q)-\mu)})$$
\begin{eqnarray}
+T\log(1+e^{-\beta(E(T,\mu,q)+\mu)})]+{M^2(T,\mu) \over \pi\lambda}.
\end{eqnarray}
For $T=0$, we approximate the effective potential as 
$$V(M,\mu,T=0)= -4\int_{0}^{\Lambda} dq[E(0,\mu,q)\Theta(E(0,\mu,q)-\mu)$$
\begin{eqnarray}
+\mu\Theta(\mu-E(0,\mu,q))]+{M^2(0,\mu) \over \pi\lambda}.
\end{eqnarray}
Integrating  Eq.(3$\cdot$24) over $q$, we get
\begin{eqnarray}
{V(M,\mu,T=0) \over \Lambda^2}=-2\left[{\bar E}(0,\mu,\Lambda)
+m^2(0,\mu)\log{1+{\bar E}(0,\mu,\Lambda)\over m(0,\mu)}\right]+{m^2(0,\mu) \over \pi\lambda}
\end{eqnarray}
for $\mu<M(0,\mu)$, and for $\mu>M(0,\mu)$, we get 
$${V(M,\mu,T=0) \over \Lambda^2}=-2\left[{\bar E}(0,\mu,\Lambda)-{\bar \Delta}{\bar \mu}+m^2(0,\mu)\log{1+{\bar E}(0,\mu,\Lambda)\over {\bar \Delta}+{\bar \mu}}\right]\Theta(1-{\bar \Delta})$$
\begin{eqnarray}
-4{\bar \mu}\left[{\bar \Delta}\Theta(1-{\bar \Delta})+\Theta({\bar \Delta}-1)\right]+{m^2(0,\mu) \over \pi\lambda}
\end{eqnarray}
 with ${\bar E}(0,\mu,\Lambda)\equiv E(0,\mu,\Lambda)/\Lambda=\sqrt{1+m^2(0,\mu)}$ and ${\bar \Delta}=\sqrt{{\bar \mu}^2-m^2(0,\mu)}$.

In Fig.6, we show the effective mass dependences of the effective potential $V(M,\mu,T=0)/\Lambda^2$ for several different values of chemical potential at $\lambda=0.3$. 

\begin{figure}
\centerline{\includegraphics[width=8cm]{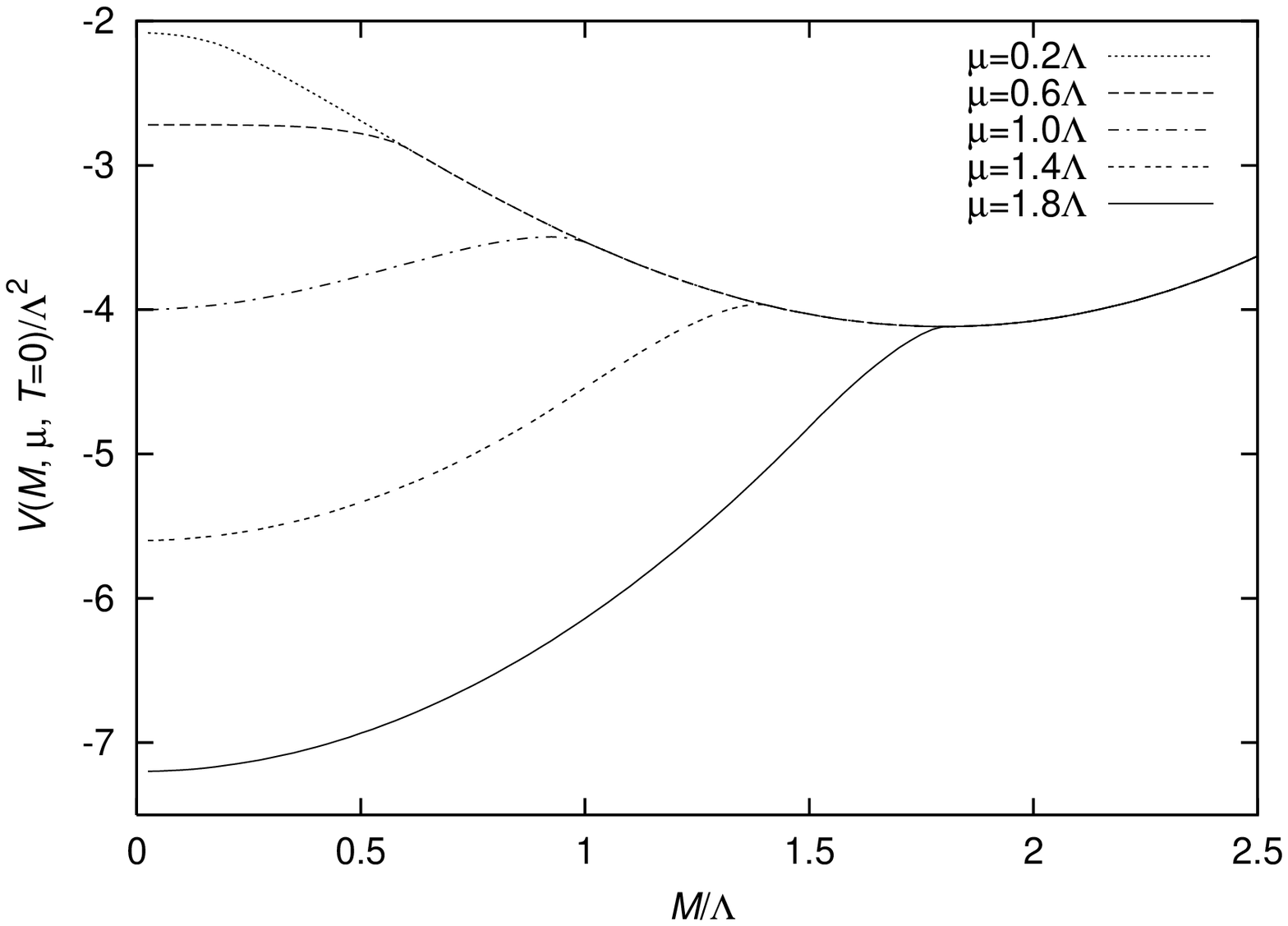}}
\caption{The effective mass dependences of the effective potential $V(M,\mu,T=0)/\Lambda^2$ for several different values of chemical potential at $\lambda=0.3$. The horizontal axis shows the effective mass divided by $\Lambda$.}
%\label{Fig.6}
\end{figure}

From Fig.6, we can see that the solutions $m^{(1)}(0,\mu)=M^{(1)}(0,\mu)/\Lambda$ and $m^{(2)}(0,\mu)=M^{(2)}(0,\mu)/\Lambda$ are the effective masses at the local minimum and maximum of the effective potential, respectively,  for ${\bar \mu}>1/\xi$ in which $m^{(2)}(0,\mu)>0$. Therefore, the SDE solution $M^{(2)}(0,\mu)$ is not a stable solution, and the stable solution is either $M=0$ or $M=M^{(1)}(0,\mu)$.

Furthermore, due to quantum effects, when $V(M=0,\mu,T=0)-V(M=M^{(1)},\mu,T=0)<0$ for ${\bar \mu}< 2\xi/(\xi^2-1)$, the solution $M^{(1)}(0,\mu)$ is a metastable solution. 

\begin{figure}
\centerline{\includegraphics[width=8cm]{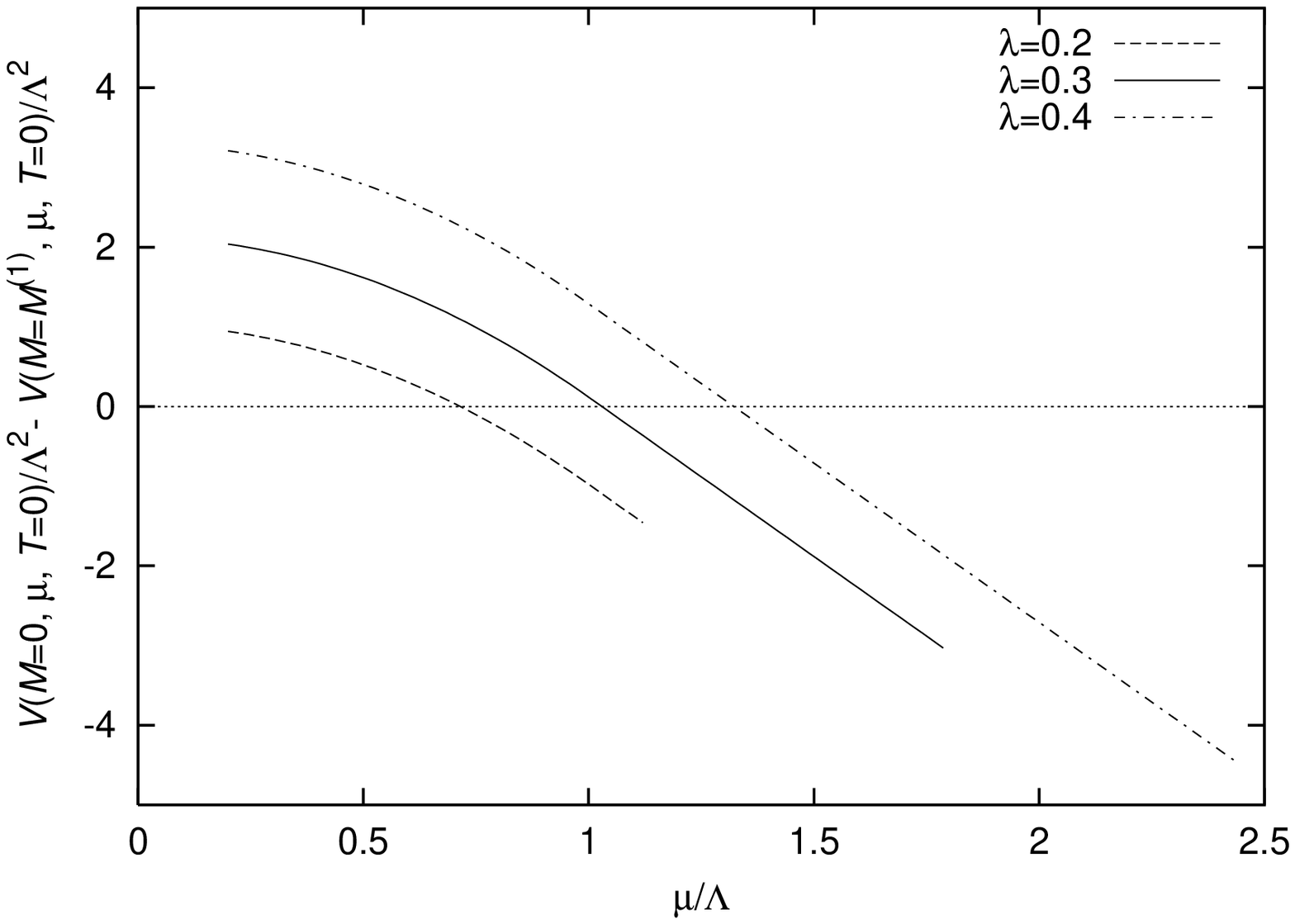}}
\caption{The chemical potential dependences of  $V(M=0,\mu,T=0)/\Lambda^2-V(M=M^{(1)},\mu,T=0)/\Lambda^2$  for several different values of $\lambda$. The horizontal axis shows the chemical potential divided by $\Lambda$. In the negative region, the solutions become metastable.}
%\label{Fig.7}
\end{figure}

As seen in Fig.7, the solution becomes metastable in the region where $\mu$ is large. Therefore, in principle, in an environment with a large chemical potential, the broken chiral symmetry tends to be restored over time due to quantum tunneling, and this situation has also been pointed out in QCD [16].

\section{Numerical solutions of the SDE for $T>0$}

In this section, we present numerical results obtained by the SDE given in Eq.(2$\cdot$9) with Eq.(2$\cdot$10) for finite temperature cases.

Fig.8 shows the effective masses solving the SDE iteratively starting from  several different initial input values $M_0$  at $T/\Lambda=0.2$ and $\lambda=0.3$.

\begin{figure}
\centerline{\includegraphics[width=8cm]{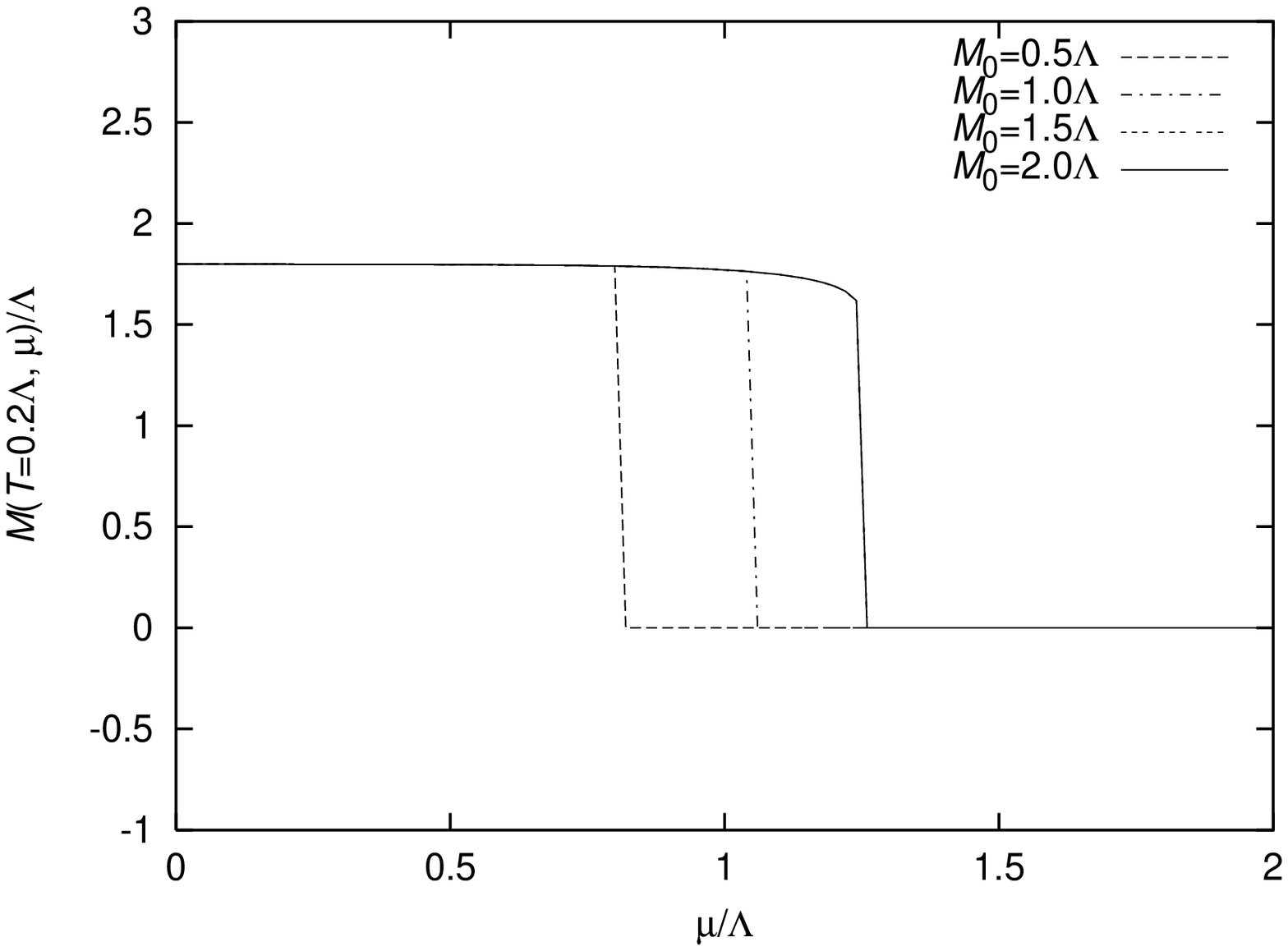}}
\caption{The chemical potential dependences  of the effective masses at $T=0.2\Lambda$ and $\lambda=0.3$ for several different initial input values $M_0$  when solving the SDE using the iterative method. The horizontal axis shows the chemical potential divided by $\Lambda$.}
%\label{Fig.8}
\end{figure}

From Fig.8, we can see that there are three  different solutions depending on the initial input values $M_0$.
Here, the solution for the initial input value $M_0=1.5\Lambda$ overlaps with the solution for $M_0=2.0\Lambda$.
 
In order to understand this behavior, we will numerically find the relation between the effective mass  and the chemical potential that satisfies the equality as
\begin{eqnarray}
1=2\pi \lambda\int_{0}^{\Lambda}  dq { 1 \over  E(T,\mu,q)} \zeta(T,\mu,E(T,\mu,q))\equiv F(M,\mu,T)
\end{eqnarray}
 with $E(T,\mu,q)=\sqrt{q^2+M^2(T,\mu)}$ and $M$ in the function $F(M, \mu, T)$ indicates $M(T,\mu)$ for given $T$ and $\mu$.
 Here, $\zeta(T,\mu,E(T,\mu,q))$ is given by Eq.(2$\cdot$10).

Fig.9 shows the effective mass dependences of $F(M,\mu,T=0.2\Lambda)$ defined in Eq.(4$\cdot$1)  for  several different values of chemical potential at $\lambda=0.3$. When changing $M$, $M$ which satisfy $F(M,\mu,T=0.2\Lambda)=1$ are the solutions of the SDE at $T=0.2\Lambda$.

\begin{figure}
\centerline{\includegraphics[width=8cm]{SDRT2MUSEARCH002B.ps}}
\caption{The effective mass dependences of $F(M,\mu,T=0.2\Lambda)$ at $\lambda=0.3$ for several different values of chemical potential. $M$ which satisfy $F(M,\mu,T=0.2\Lambda)=1$ are the solutions of the SDE at $T=0.2\Lambda$.
 The horizontal axis shows the effective mass divided by $\Lambda$.}
%\label{Fig.9}
\end{figure}

In Fig.9, except for the case of $\mu=0.6\Lambda$,  there are two solutions that satisfy $F(M, \mu, T=0.2\Lambda)=1$ for each $\mu$. The solution with the larger value of $M/\Lambda$ that satisfies $F(M,\mu,T=0.2\Lambda)=1$ is the effective mass at the local minimum of the effective potential $V(M,\mu,T=0.2\Lambda)$ and corresponds to the solution $m^{(1)}(T=0.2\Lambda,\mu)=M^{(1)}(T=0.2\Lambda,\mu)/\Lambda$. On the other hand, the solution with the smaller value of $M/\Lambda$ that satisfies $F(M,\mu,T=0.2\Lambda)=1$ is the effective mass at the local maximum of the effective potential $V(M,\mu,T=0.2\Lambda)$ and corresponds to the solution $m^{(2)}(T=0.2\Lambda,\mu)=M^{(2)}(T=0.2\Lambda,\mu)/\Lambda$.

To see the situation mentioned above, Fig.10 shows the effective mass dependences of the effective potential $V(M,\mu,T=0.2\Lambda)$ for  several different values of chemical potential at $\lambda=0.3$. Here, the effective potential $V(M, \mu, T)$ is given by Eq.(3$\cdot$23).

\begin{figure}
\centerline{\includegraphics[width=8cm]{EFTVMT2MU003B.ps}}
\caption{The effective mass dependences of the effective potential $V(M,\mu,T=0.2\Lambda)/\Lambda^2$ for several different values of chemical potential at  $\lambda=0.3$. The horizontal axis shows the effective mass divided by $\Lambda$.}
%\label{Fig.10}
\end{figure}
From Fig.10, we can see that no local maximum is found in the effective mass dependence of the effective potential at $\mu=0.6\Lambda$.

Fig.11 shows the chemical potential dependences of the solutions that satisfy Eq.(4$\cdot$1)  at $T/\Lambda=0.2$ and $\lambda=0.3$.
 The solution $M^{(1)}(T=0.2\Lambda,\mu)/\Lambda$ 
 is denoted by the solid curve and  the solution $M^{(2)}(T=0.2\Lambda,\mu)/\Lambda$ is denoted by the dashed-dotted curve.
 
\begin{figure}
\centerline{\includegraphics[width=8cm]{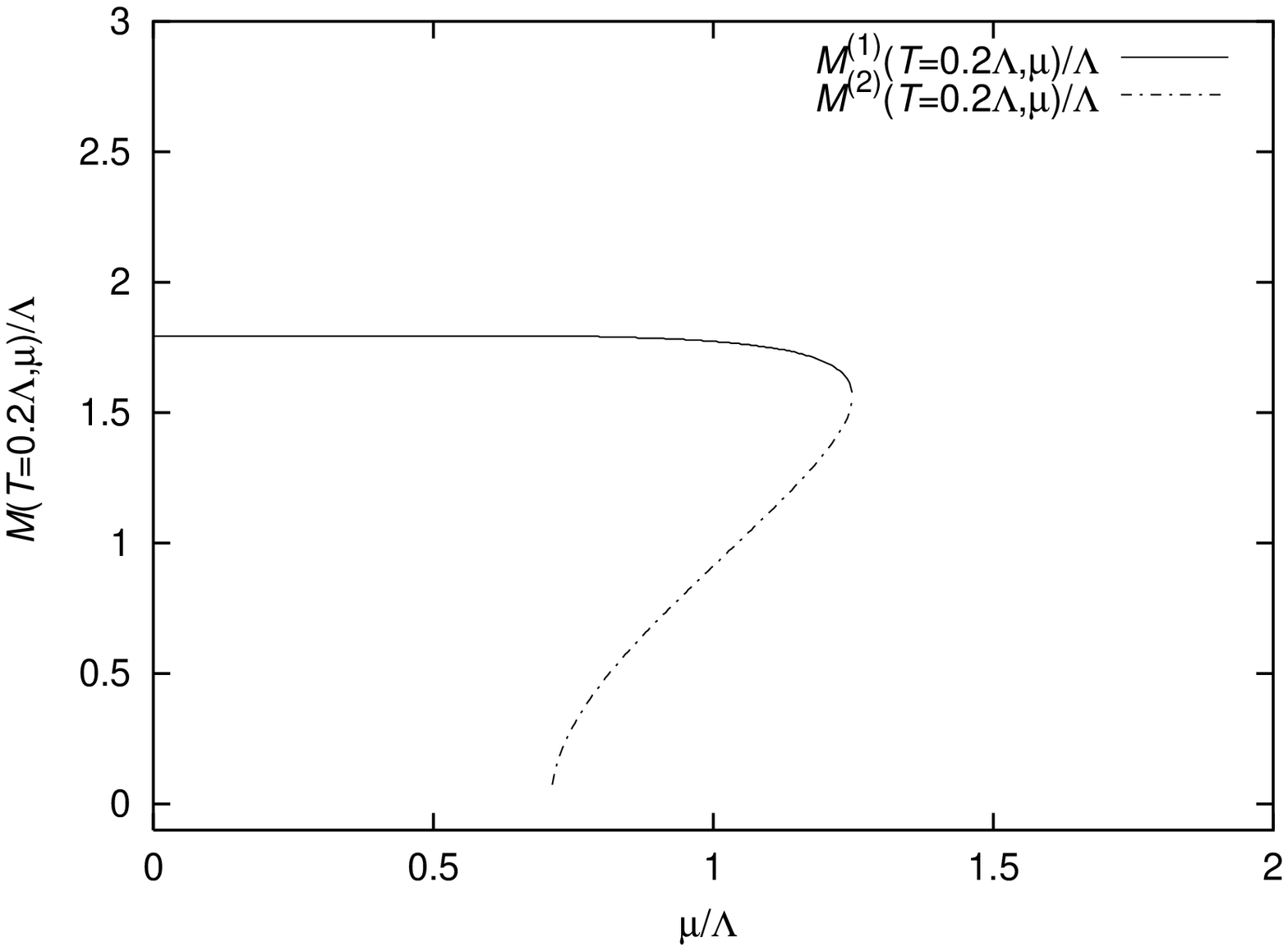}}
\caption{The chemical potential dependences of the solutions that satisfy Eq.(4$\cdot$1) at $T/\Lambda=0.2$ and $\lambda=0.3$. The solutions $M^{(1)}(T=0.2\Lambda,\mu)/\Lambda$ and $M^{(2)}(T=0.2\Lambda,\mu)/\Lambda$ are denoted by the solid curve and the dashed-dotted curve, respectively. The horizontal axis shows the chemical potential divided by $\Lambda$.}
%\label{Fig.11}
\end{figure}

 Now, let us denote the value of the chemical potential at the end point of the solution $M^{(1)}(T=0.2\Lambda,\mu)$ in Fig. 11 as $\mu_{\rm C}$. 
 Then, we can see that $M^{(1)}(T=0.2\Lambda,\mu_{\rm C}) \simeq M^{(2)}(T=0.2\Lambda,\mu_{\rm C})\simeq 1.5\Lambda$ holds at $\mu_{\rm C}$.
 Here, $\mu_{\rm C}$ corresponds to the chemical potential at the phase transition point (critical chemical potential).
Therefore, in Fig.8, when solving the SDE using the iterative method, the solutions with the initial input values $M_0=1,5\Lambda$ and $M_0=2.0\Lambda$
 become zero at almost the same value of the chemical potential $\mu_{\rm C}$.
 
Next, Fig.12 shows the case where the SDE is solved by the iterative method at $T/\Lambda=0.4$  and $\lambda=0.3$.
\begin{figure}
\centerline{\includegraphics[width=8cm]{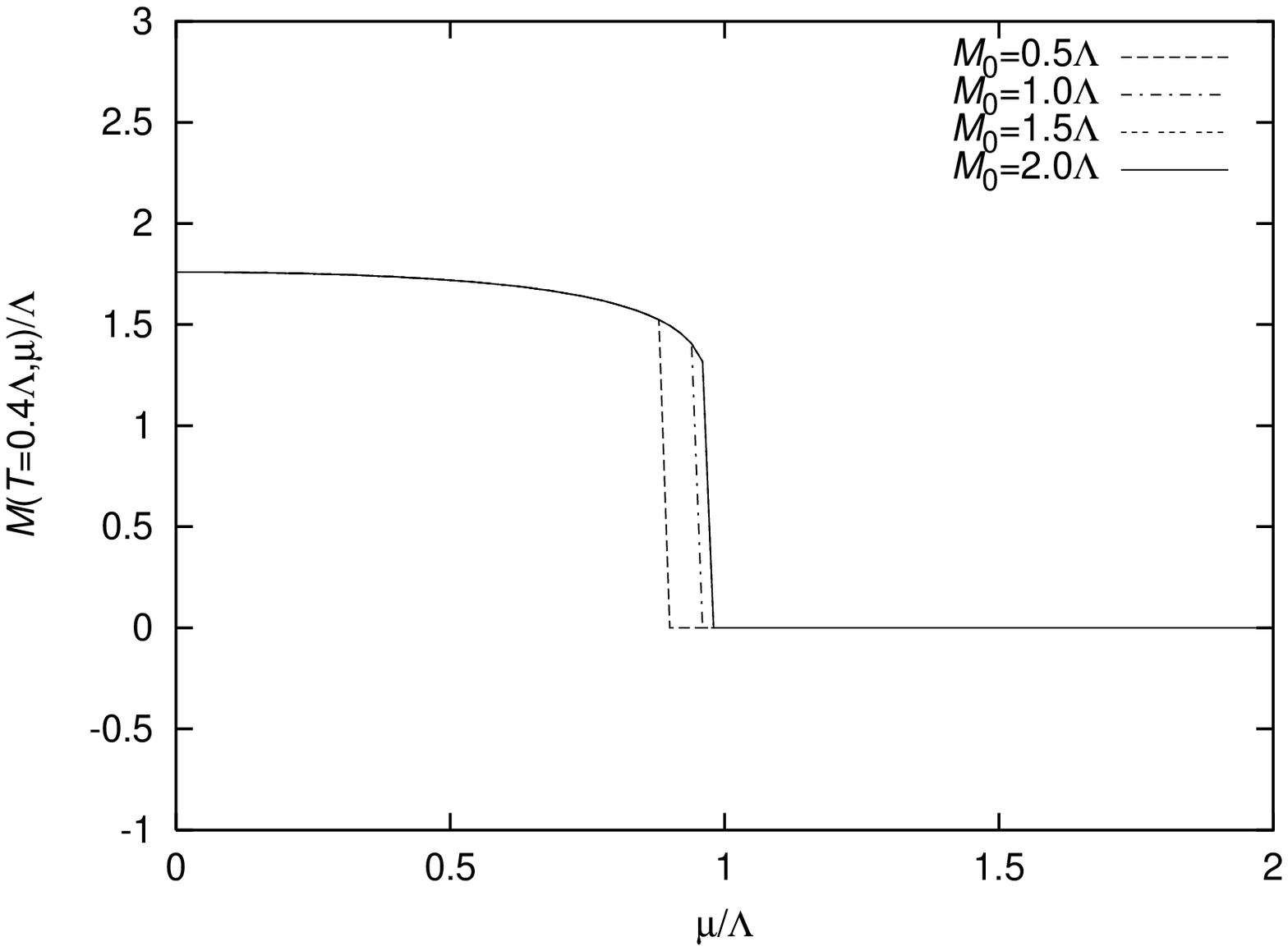}}
\caption{The chemical potential dependences  of the effective masses at $T/\Lambda=0.4$ and $\lambda=0.3$ for  several different initial input values $M_0$  when solving the SDE using the iterative method. The horizontal axis shows the chemical potential divided by $\Lambda$.}
%\label{Fig.12}
\end{figure}
 In Fig.12, the solution for the initial input value $M_0=1.5\Lambda$ overlaps with the solution for $M_0=2.0\Lambda$.
Note that  the differences among  the solutions  for different values of $M_0$ are smaller than in the case of $T/\Lambda=0.2$ shown in Fig.8.
\begin{figure}
\centerline{\includegraphics[width=8cm]{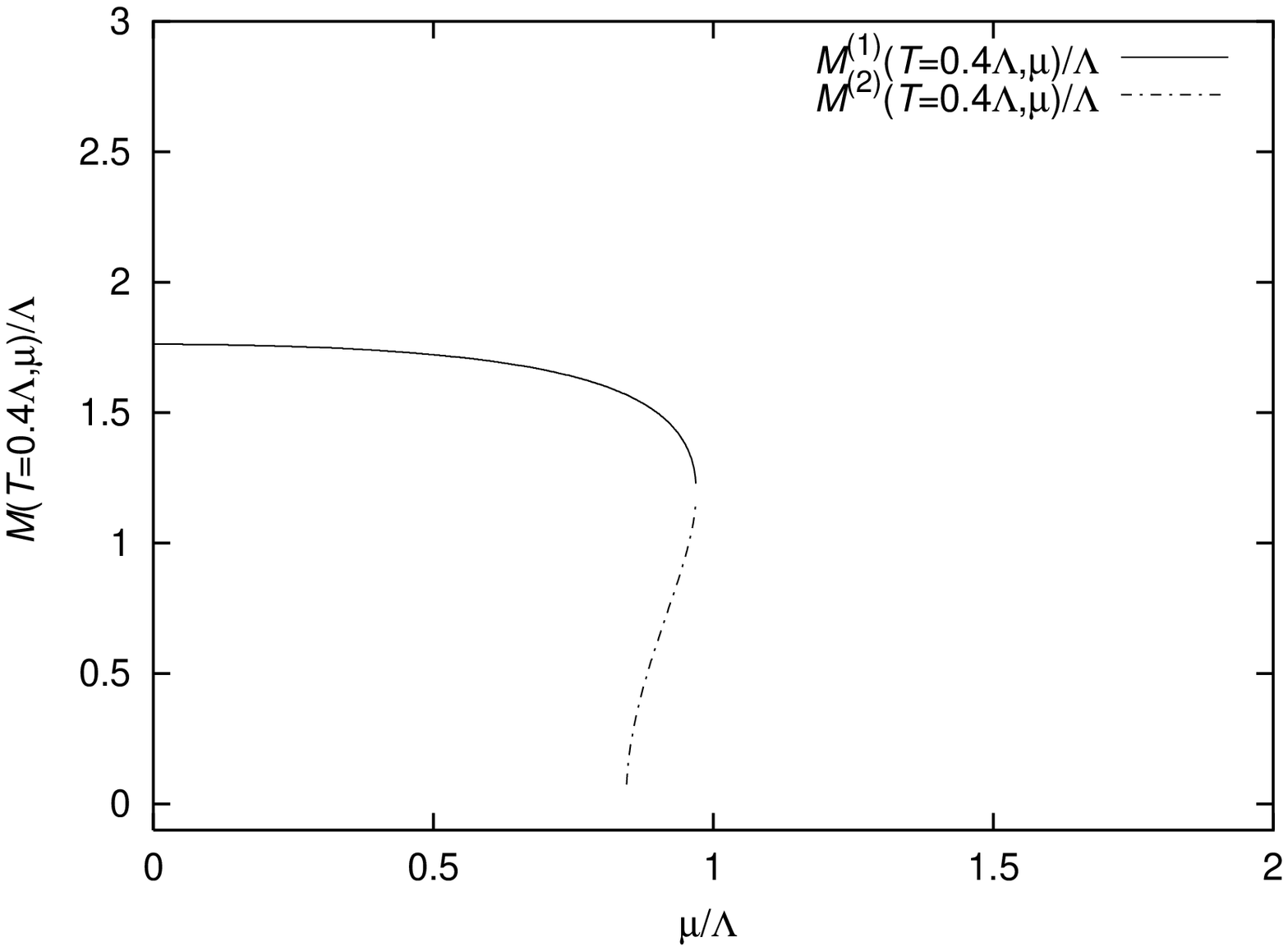}}
\caption{The chemical potential dependences of the solutions that satisfy Eq.(4$\cdot$1) at $T/\Lambda=0.4$  and $\lambda=0.3$. The solutions $M^{(1)}(T=0.4\Lambda,\mu)/\Lambda$ and $M^{(2)}(T=0.4\Lambda,\mu)/\Lambda$ are denoted by the solid curve and the dashed-dotted curve, respectively. The horizontal axis shows the chemical potential divided by $\Lambda$.}
%\label{Fig.13}
\end{figure}

 Fig.13 shows the chemical potential dependences of the solutions that satisfy Eq.(4$\cdot$1) at $T/\Lambda=0.4$ and $\lambda=0.3$.
 The solution $M^{(1)}(T=0.4\Lambda,\mu)/\Lambda$ is denoted by the solid curve and the solution $M^{(2)}(T=0.4\Lambda,\mu)/\Lambda$ is denoted by the dashed-dotted curve.
  Comparing Fig.13 and Fig.11, the range of the chemical potential  for the solution $M^{(2)}(T=0.4\Lambda,\mu)/\Lambda$ gathers around the critical chemical potential (the end point of the solution $M^{(1)}(T=0.4\Lambda,\mu)/\Lambda$).

 Fig.14 shows the case where the SDE is solved by the iterative method at $T/\Lambda=0.6$  and $\lambda=0.3$. In Fig.14, it can be seen that the solutions for all $M_0$ values overlap.
\begin{figure}
\centerline{\includegraphics[width=8cm]{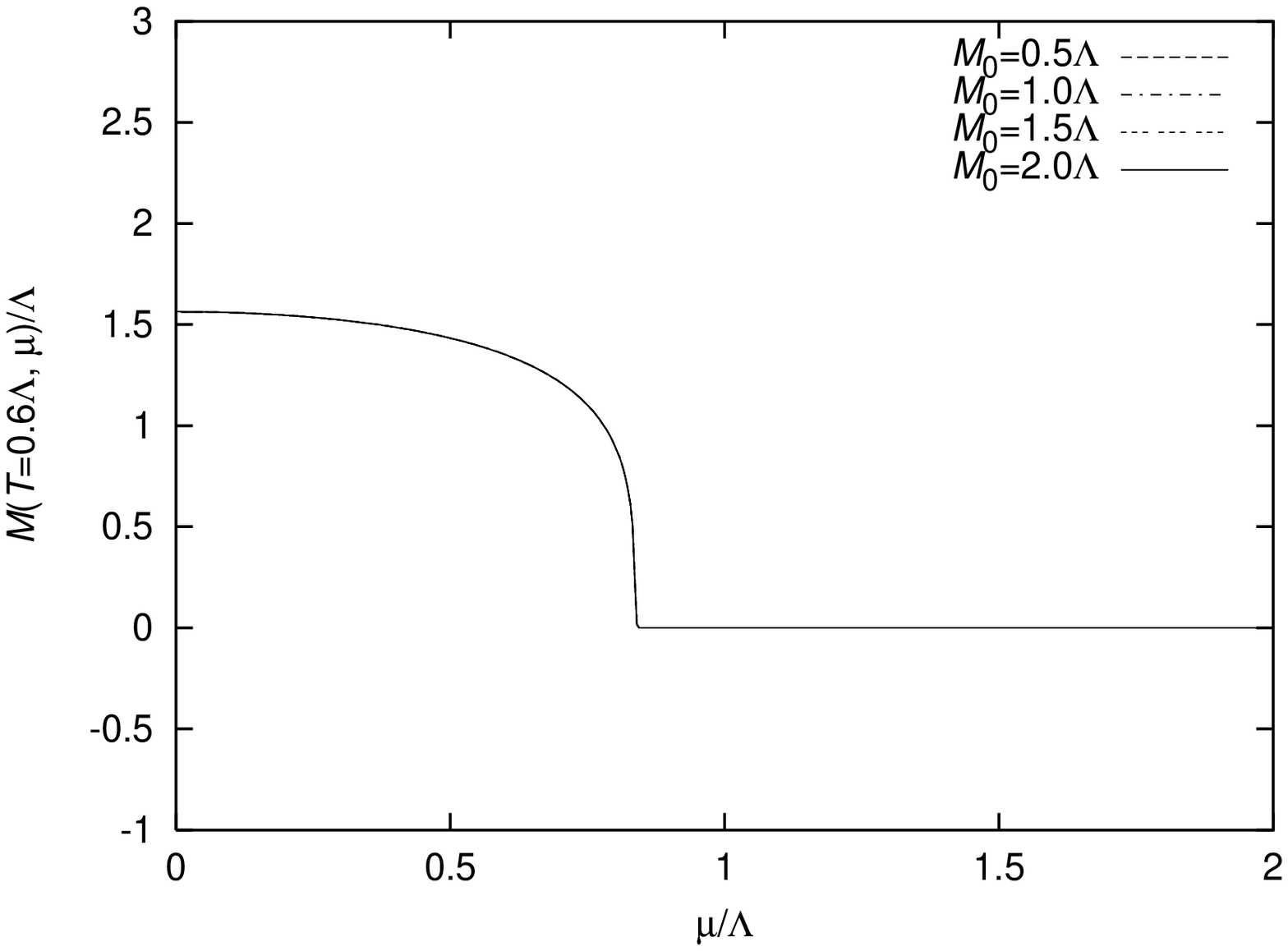}}
\caption{The chemical potential dependences of the effective masses at $T/\Lambda=0.6$  and $\lambda=0.3$ for several different initial input values $M_0$  when solving the SDE using the iterative method. The horizontal axis shows the chemical potential divided by $\Lambda$.}
%\label{Fig.14}
\end{figure}
\begin{figure}
\centerline{\includegraphics[width=8cm]{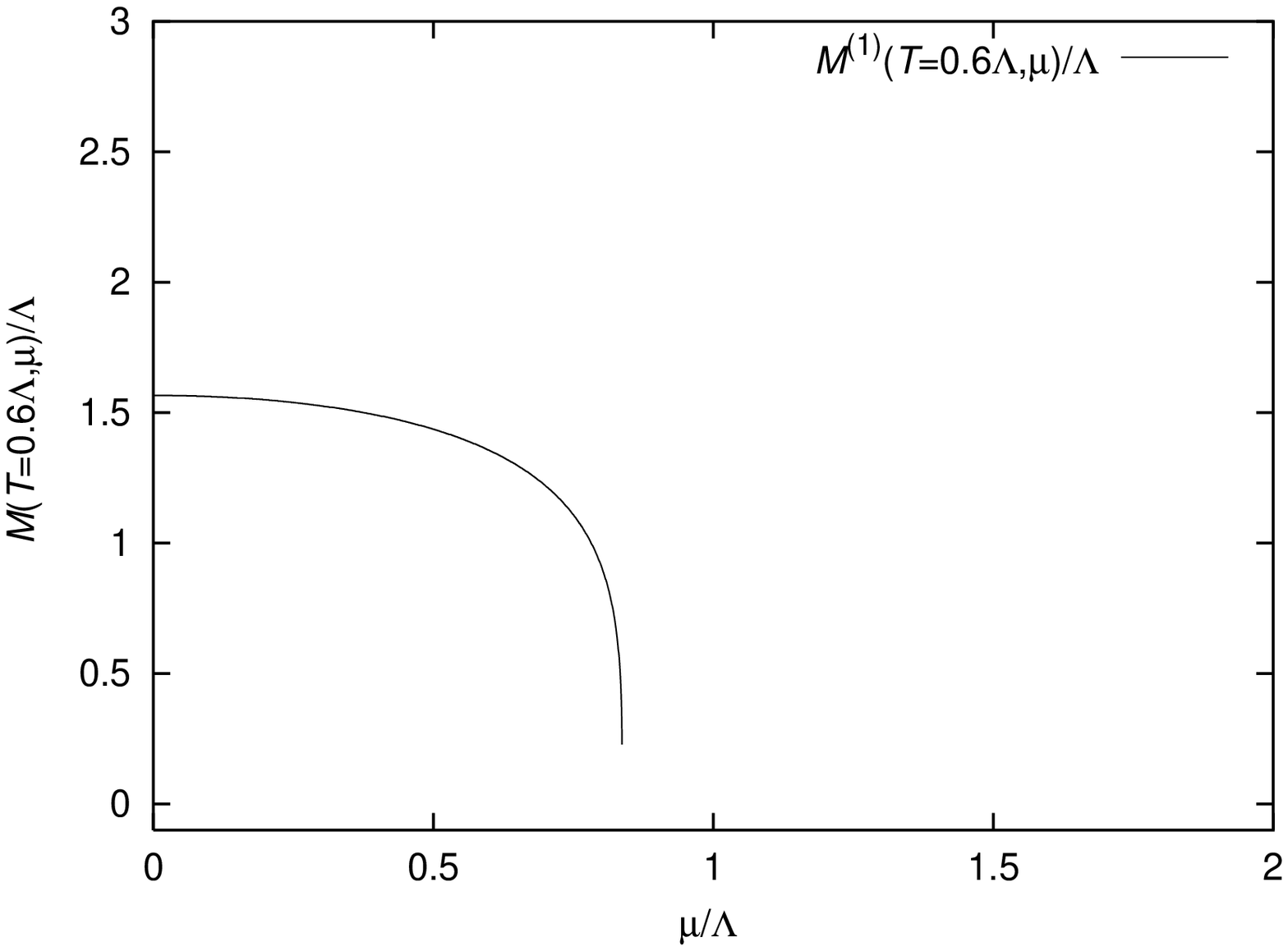}}
\caption{The chemical potential dependence of the solution that satisfies Eq.(4$\cdot$1) at $T/\Lambda=0.6$ and $\lambda=0.3$. The solution $M^{(1)}(T=0.6\Lambda,\mu)/\Lambda$ is denoted by the solid curve.  The horizontal axis shows the chemical potential divided by $\Lambda$.}
%\label{Fig.15}
\end{figure}

Fig.15 shows the chemical potential dependence of the solution that satisfies 
 Eq.(4$\cdot$1) at $T/\Lambda=0.6$ and $\lambda=0.3$.
 As shown in Fig.15, the solution $M^{(2)}(T=0.6\Lambda,\mu)$ disappears at $T/\Lambda=0.6$.
Furthermore, this area is the boundary of the temperature between the first-order phase transition and the second-order phase transition along the chemical potential axis. This observation suggests that the existence of the solution $M^{(2)}(T,\mu)$ is related to whether the phase transition is first-order or second-order.

Fig.16 shows the effective mass dependences of $F(M,\mu,T=0.6\Lambda)$ for $\mu$ near the critical chemical potential at $\lambda=0.3$.
\begin{figure}
\centerline{\includegraphics[width=8cm]{SDRT6MU8SEARCH003B.ps}}
\caption{The effective mass dependences of $F(M,\mu,T=0.6\Lambda)$ at $\lambda=0.3$ for $\mu$ near the critical chemical potential. $M$ which satisfy $F(M,\mu,T=0.6\Lambda)=1$ are the solutions of the SDE at $T=0.6\Lambda$.
 The horizontal axis shows the effective mass divided by $\Lambda$.}
%\label{Fig.16}
\end{figure}
As can be seen from Fig.16, the effective mass dependence of $F(M,\mu,T)$ is flatter than at low temperatures. As $\mu$ approaches the critical chemical potential, the solution $M^{(1)}(T,\mu)$ gradually approaches zero, but there is no point of intersection with $F(M,\mu,T)=1$ corresponding to the solution $M^{(2)}(T,\mu)$ near the critical chemical potential.

In general, we find that the effective mass dependences of $F(M,\mu,T)$ are flatter at higher temperatures than at lower temperatures.
As examples, Figs.17 and 18 show the effective mass dependences  of $F(M,\mu,T)$ and the effective potential, respectively, at $\mu=0.9\Lambda$ and $\lambda=0.3$.
\begin{figure}
\centerline{\includegraphics[width=8cm]{SDRTMU9SEARCH001B.ps}}
\caption{The effective mass dependences of $F(M,\mu=0.9\Lambda,T)$ at $\lambda=0.3$ for several different values of temperature. $M$ which satisfy $F(M,\mu=0.9\Lambda,T)=1$ are the solutions of the SDE at $\mu=0.9\Lambda$.
 The horizontal axis shows the effective mass divided by $\Lambda$.}
%\label{Fig.17}
\end{figure}

\begin{figure}
\centerline{\includegraphics[width=8cm]{EFTVMMU9T001B.ps}}
\caption{The effective mass dependences of the effective potential $V(M,\mu=0.9\Lambda,T)/\Lambda^2$ for several different values of temperature at $\lambda=0.3$.  The horizontal axis shows the effective mass divided by $\Lambda$.}
%\label{Fig.18}
\end{figure}
From Fig.17, we see that as the temperature increases, the dependence of $F(M,\mu,T)$ on the effective mass becomes monotonic and less sensitive to the chemical potential. This behavior may be understood as the effect of the weakening of the dependence of the Boltzmann factor $e^{\beta(E-\mu)}$ as $\beta=1/T$ becomes smaller. In Fig.17, as the temperature increases, the distribution of $F(M,\mu,T)$ gradually flattens out, and both solutions $M^{(1)}$ and $M^{(2)}$ move to smaller regions. Eventually, $F(M,\mu,T)=1$ is no longer satisfied.

For example, at the point $T=0.6\Lambda$ and $\mu=0.9\Lambda$, $F(M,\mu,T)\neq 1$ is satisfied for all effective masses, and the only solution to the SDE given by $M=MF(M,\mu,T)$ is $M=0$. As can be seen in Fig.18, the effective potential also has a flat dependence on the effective mass as the temperature increases.
This result shows that the dependence of the solution obtained by the iterative method on the initial input value disappears at high temperatures.

On the other hand, at low temperatures, the effect of the chemical potential is strongly evident in the $M$ dependence of $F(M,\mu,T)$. This behavior may be understood as the effect of the fermion occupancy rate changing significantly at the Fermi energy. At low temperatures, the $M<\mu$ region contains contributions of energies smaller than the Fermi energy, while the $M>\mu$ region contains contributions mainly of energies larger than the Fermi energy. Since the energy depends on the effective mass, the effect of the chemical potential may be understood as appearing in the $M$ dependence of $F(M,\mu,T)$. However, this behavior is suppressed as temperature increases due to the properties of the Boltzmann factor.

\section{How to avoid the effects of spurious solution}

In this section, based on the analyses in the previous sections, we consider how to select initial input values to  avoid  spurious solutions when using the iterative method.

Let us summarize what can be said from the analyses so far about selecting initial values at a given temperature and density.

First, from Figs.2,11,13,and 15,  it seems that there is no solution $M^{(2)}(T,\mu)$ for $\mu/\Lambda<1/\xi=e^{-1/(2\pi\lambda)}$ at any $T$.
In this region, the effective mass $M$ dependence of the effective potential $V$ does not have a negative curvature at $\partial V/\partial M=0$, which means that the effective potential does not have a local maximum due to the change in the effective mass.  Therefore, there is no need to restrict the initial input value in this region.
Also, at $T=0$, $M^{(2)}(0,\mu)$ satisfies $M^{(2)}(0,\mu)\leq \mu$.(See Fig.2.)
As mentioned in Section 3.3, if the initial input value $M_0$ is set to $M_0>M^{(2)}(0,\mu)$, the solution obtained by the iterative method will converge to $M^{(1)}(0,\mu)$. In this case, the effects of spurious solution does not appear. Therefore, if we select the initial input value as $M_0>\mu$, the effects of spurious solution can be prevented. The above properties hold for any value of $\lambda$.

However, the choice of $M_0>\mu$ for the initial input value does not necessarily hold for $T>0$.
Fig.19 shows the chemical potential dependences of $M^{(2)}(T,\mu)$ divided by $\mu$ for several different temperatures $T$ at $\lambda=0.3$.
\begin{figure}
\centerline{\includegraphics[width=8cm]{SERCHRM2DIVMUVSMU001B.ps}}
\caption{The chemical potential dependences  of $M^{(2)}(T,\mu)$ divided by $\mu$ for several different temperatures $T$ at $\lambda=0.3$. The horizontal axis shows the chemical potential divided by $\Lambda$.}
%\label{Fig.19}
\end{figure}
As can be seen from Fig.19, there exists a region of $\mu$ such that $M^{(2)}(T,\mu)>\mu$ for $T>0$. Therefore, further consideration is required to select the initial input values at finite temperatures.

As we show in Appendix A, in our model $M^{(1)}(0,0)(=M^{(1)}(0,\mu))$ is analytically calculable and is the maximal solution of the SDE.
 $M^{(1)}(0,0)$ can be calculated from Eq.(3$\cdot$7) as $M^{(1)}(0,0)/\Lambda=2\xi/(\xi^2-1)$. Therefore, since $M^{(1)}(0,0)>M^{(1)}(T>0,\mu)$ and $M^{(1)}(T,\mu)\geq M^{(2)}(T,\mu)$, a simple candidate for the initial input value $M_0$ to obtain the effective mass at finite temperature and finite density seems to be $M_0 = M^{(1)}(0,0)$. (For more details, see Appendix A.) 
 As mentioned earlier, the solution $M^{(1)}(0,0)$ can also be found using an iterative method with any initial input value.

 What needs to be checked is whether there is only one solution that corresponds to a local minimum of the effective potential.
 As shown in the previous sections, our model has a unique solution $M^{(1)}(T,\mu)$ corresponding to a local minimum of the effective potential for each $T$ and $\mu$.\footnote{See Appendix B for a method to use an iterative method to check whether there are multiple solutions corresponding to local minima in the effective potential.} 
 
 When there is only one solution $M^{(1)}(T,\mu)$ for each $T$ and $\mu$, the following method may be used.
The first step is to solve the solution for $T=\mu=0$  using an iterative method with any initial input value or an analytical solution and use this value as the initial input for estimating the effective mass in $M^{(1)}(T+\Delta T,\mu+\Delta\mu)$.[17]  For this method to be effective, it is necessary to ensure that the effective masses $M$ during the iterations are never smaller than the solution $M^{(2)}(T+\Delta T,\mu+\Delta\mu)$, even for sufficiently small intervals $\Delta T$ and $\Delta\mu$. The safest situation may appear to be when the effective mass $M$ calculated by the SDE converges from a value larger than $M^{(1)}(T+\Delta T,\mu+\Delta\mu)$ to $M^{(1)}(T+\Delta T,\mu+\Delta\mu)$. Our model may appear to satisfy this situation.(See Appendix B for details.)

\section{Summary and Comments}

When evaluating finite temperature and/or density systems,
 the Schwinger-Dyson equation (SDE) is often used to find non-perturbative solutions.
In this case, if the SDE cannot be solved analytically, a method is used to find  the solutions by iteratively using the SDE with an initial input value as a effective mass. 

 In this paper, we investigated the initial input value dependence that appears when solving the SDE using the iterative method.
 As a simple example, we solved the SDE for a model with four-fermion interactions in the (1+1) space-time dimensions.  In this model, there is a solution that breaks chiral symmetry, and massless fermions acquire effective mass, but this symmetry is restored at high temperatures or high densities.

 The solutions obtained by the SDE correspond to the effective masses at the extreme values of the effective potential. In our case, we showed that  for each value of temperature $T$ and chemical potential $\mu$ there are regions with two solutions corresponding to the effective masses at the local minimum and local maximum values of the effective potential. We denoted these solutions  as $M^{(1)}(T,\mu)$ and $M^{(2)}(T,\mu)$, respectively.
 The spurious solution $M^{(2)}(T,\mu)$, which corresponds to the effective mass at a local maximum of the effective potential, is unstable and is usually excluded.
  However, when solving the SDE using the iterative method, this solution does not clearly appear. Instead, it appears as dependences of the solutions on the initial input values and they give false phase transition solutions.

 To understand this situation, we first examined the properties of the solution of the SDE at zero temperature and finite density. 
Since the model we dealt with can be solved analytically at the low temperature limit, we compared the analytical solutions with the numerical results obtained by the iterative method and discussed the relation between them. 

  Knowing the properties of the solutions investigated in this paper,  there may be several ways to avoid the spurious solutions  in various ranges of temperatures and densities.
 First, we found that the spurious  solutions do not appear in the region where the chemical potential $\mu$ satisfies
$\mu/\Lambda< e^{-1/(2\pi\lambda)}$ at any temperature $T$, where $\Lambda$ and $\lambda$ denote a cutoff in the momentum integral of the SDE and coupling constant, respectively. Therefore, there are no restrictions on the initial input value when the chemical potential is in this range.
 In the case of zero temperature and finite density for $\mu/\Lambda \geq e^{-1/(2\pi\lambda)}$ , the spurious solution $M^{(2)}(0,\mu)$ is equal to or smaller than the chemical potential $\mu$, so the appearance of the spurious solution can be prevented by choosing the initial input value $M_0$ as $M_0>\mu$. Starting from this initial input, the solution of the SDE converges to $M^{(1)}(0,\mu)$.
  However, at finite temperatures there exists a region where the restriction $M_0>\mu$ is not necessarily valid. Since we found the relation $M^{(1)}(0,0)>M^{(1)}(T>0,\mu)$, we may choose $M^{(1)}(0,0)$ as the initial input value at finite temperature. The solution $M^{(1)}(0,0)$ is obtained by iteratively solving the SDE at $T=\mu=0$. In this case, the solution does not depend on the initial input values. In our model, $M^{(1)}(0,0)$ can also be obtained analytically. 
 
 Additionally, we have discussed the use of the solution $M^{(1)}(0,0)$ as the initial input for evaluating $M^{(1)}(T+\Delta T,\mu+\Delta\mu)$ [17]. It is pointed out that when using this method, it is necessary to check the existence of multiple solutions and the convergence properties of the solutions. (See Appendix B.)
 
 Our study suggests that when solving nonlinear equations numerically, it is necessary to carefully examine the parameter space of initial input values and choose suitable initial values.
 
Although the problem investigated in this paper using a simple model is technical problems when solving nontrivial solutions numerically using nonlinear equations such as the SDE, these results may also provide useful information when solving more complicated systems.

\vspace{5mm}

\vspace{5mm}
\begin{center}
{\large \bf Appendix A. Properties of the SDE solution}
\end{center}

\vspace{2mm}

In this appendix, we further explore the properties of the solutions obtained by the SDE and consider how to find correct solutions for finite temperature $T$ and chemical potential $\mu$ while avoiding spurious solutions when solving the SDE iteratively.

From Eqs.(2$\cdot$9) and (2$\cdot$10), the SDE for the effective mass $M(T,\mu)$ is given as
$$M(T,\mu)=M(T,\mu)F(M,\mu,T),$$
with
$$F(M,\mu,T)=2\pi\lambda\int_{0}^{\Lambda} dq{1 \over E(T,\mu,q)}\zeta(T,\mu,E(T,\mu,q)).$$
Here, $\zeta(T,\mu,E(T,\mu,q))$ given in Eq.(2$\cdot$10)  can be written as 
$$\zeta(T,\mu,E(T,\mu,q))=1-{1 \over 1+e^{\beta(E(T,\mu,q)-\mu)}}-{1 \over 1+e^{\beta(E(T,\mu,q)+\mu)}}.$$

The SDE is given by the extremum 
$${\partial V \over \partial M}=-4M(T,\mu)F_0(M,\mu,T)+{2M(T,\mu) \over \pi\lambda}=0$$ 
of the effective potential $V$ given by Eq.(3$\cdot$23) with respect to the effective mass $M(T,\mu)$. Here, $F_0(M,\mu,T)$ is defined as
$F(M,\mu,T)=2\pi\lambda F_0(M,\mu,T)$.
 From the above equation, $\partial V / \partial M>0$, $\partial V / \partial M<0$ and $\partial V / \partial M=0$ correspond to $F(M,\mu,T)<1$, $F(M,\mu,T)>1$ and $F(M,\mu,T)=1$, respectively, for $M(T,\mu)\neq 0$.

Now, if we write $M^{(1)}(T,\mu)$ as the effective mass at a local minimum of the effective potential, we can see that for $M^{(1)}(T,\mu)>M$, $\partial V/\partial M<0$ and $F(M.\mu,T)>0$ are satisfied. Conversely, for $M^{(1)}(T,\mu)<M$, we can see that $\partial V/\partial M>0$ and $F(M.\mu,T)<0$ are satisfied.

On the other hand, if we write the effective mass at the local maximum of the effective potential as $M^{(2)}(T,\mu)$, we can see that for $M^{(2)}(T,\mu)>M$, $\partial V/\partial M>0$ and $F(M.\mu,T)<0$ are satisfied. Conversely, for $M^{(2)}(T,\mu)<M$, we can see that $\partial V/\partial M<0$ and $F(M.\mu,T)>0$ are satisfied.

Therefore, by examining the sign of $\partial V/\partial M$ or $F(M.\mu,T)-1$ near $\partial V/\partial M=0$ or $F(M.\mu,T)=1$, it is possible to determine whether the solution is at an unstable local maximum or a stable local minimum.

Here, $F(M,\mu,T)=1$ can be written as
$$ \int_{0}^{\Lambda} dq{1 \over E(T,\mu,q)}-{1 \over 2\pi\lambda}
=\int_{0}^{\Lambda} dq{1 \over E(T,\mu,q)}\left[1/(1+e^{\beta(E(T,\mu,q)-\mu)})+1/(1+e^{\beta(E(T,\mu,q)+\mu)})\right].~~~({\rm A}\cdot 1)$$

At $T=\mu=0$, since the right side of the above equation vanishes,  Eq.(A$\cdot$1) becomes
$$ \int_{0}^{\Lambda} dq{1 \over E(0,0,q)}-{1 \over 2\pi\lambda}
=\log{\Lambda+\sqrt{\Lambda^2+M^2(0,0)} \over M(0,0)}-{1 \over 2\pi\lambda}
=0.$$
From this equation, the non-trivial solution for $T=\mu=0$ is given by
$$ M(0,0)/\Lambda={2\xi \over \xi^2-1}$$
with $\xi=e^{1/(2\pi\lambda)}$, which corresponds to the case with $\mu=0$ in the solution $M^{(1)}(0,\mu)/\Lambda$ given in Eq.(3$\cdot$7).

In addition, since  the first term on the left side of Eq.(A$\cdot$1) decreases as the effective mass increases, if there is a solution that satisfies $M(T>0,\mu)>M^{(1)}(0,0)$, we have
$$\int_{0}^{\Lambda} dq{1 \over E(T>0,\mu,q)}-{1 \over 2\pi\lambda}=\log{\Lambda+\sqrt{\Lambda^2+M^2(T>0,\mu)} \over M(T>0,\mu)}-{1 \over 2\pi\lambda}<0.$$
On the other hand, since the right side of Eq.(A$\cdot$1) is positive for finite $T$, there is no solution that satisfies $M(T>0,\mu)>M^{(1)}(0,0)$. Therefore, the solution at finite temperature satisfies $M^{(1)}(T>0,\mu)<M^{(1)}(0,0)$.

Next, by  differentiating $F(M,\mu,T)$ with respect to the effective mass $M$, we get
$$F'(M,\mu,T)\equiv {\partial F \over \partial M}=-2\pi\lambda\int_{0}^{\Lambda} dq{M(T,\mu) \over E^3(T,\mu,q)}$$
$$+2\pi\lambda\int_{0}^{\Lambda} dq{M(T,\mu) \over E(T,\mu,q)}\left\{{-\zeta(T,\mu,E(T,\mu,q))+1 \over E^2(T,\mu,q)}+{1 \over E(T,\mu,q)}{\partial \zeta \over \partial E}\right\}~~~({\rm A}\cdot 2)$$
with
$${\partial \zeta \over \partial E}={\beta e^{\beta(E(T,\mu,q)-\mu)} \over (1+e^{\beta(E(T,\mu,q)-\mu)})^2}+{\beta e^{\beta(E(T,\mu,q)+\mu)} \over (1+e^{\beta(E(T,\mu,q)+\mu)})^2}$$
and
$$-\zeta(T,\mu,E(T,\mu,q))+1=1/(1+e^{\beta(E(T,\mu,q)-\mu)})+1/(1+e^{\beta(E(T,\mu,q)+\mu)}).$$
Here, the first term of Eq.(A$\cdot$2) has a negative value.
 At $T=\mu=0$, the second term of Eq.(A$\cdot$2) vanishes.
 Since the second term in Eq.(A$\cdot$2) is positive when $T>0$, at finite temperatures we have $F'(M,\mu,T>0)>F'(M,0,0)$.
  Since $F'(M,0,0)<0$, $F(M,0,0)$ is a decreasing function with respect to $M$.
 Furthermore, since $F(M,0,0)>1$ for small $M(0,0)$, there is only one solution that satisfies $F(M,0,0)=1$.
 In Fig. 20 and Fig. 21, we show the effective mass dependences for several different temperatures of $F(M,\mu=0,T)$ and the effective potential $V(M,\mu=0,T)$ , respectively, at $\lambda=0.3$.
 \begin{figure}
\centerline{\includegraphics[width=8cm]{SDRTMU0SEARCH001B.ps}}
\caption{The effective mass dependences of $F(M,\mu=0,T)$ at $\lambda=0.3$ for several different values of temperature. $M$ which satisfy $F(M,\mu=0,T)=1$ are the solutions of the SDE at $\mu=0$.
 The horizontal axis shows the effective mass divided by $\Lambda$.}
%\label{Fig.20}
\end{figure}

 \begin{figure}
\centerline{\includegraphics[width=8cm]{EFTVMMU0T001B.ps}}
\caption{The effective mass dependences of the effective potential $V(M,\mu=0,T)/\Lambda^2$ for several different values of temperature at $\lambda=0.3$. The horizontal axis shows the effective mass divided by $\Lambda$.}
%\label{Fig.21}
\end{figure}
 
 In the vicinity of the solution $M^{(1)}(0,0)$, $F(M_0,0,0)<1$ for the initial value $M_0>M^{(1)}(0,0)$. Therefore, by using the SDE, we obtain $M_1=M_0F(M_0,0,0)<M_0$. If we continue to perform iterative calculations using the SDE, the effective mass decreases as $M_1>M_2>\cdots$ and converges when $F(M,0,0)=1$. \footnote{If it does not converge and there is a slight deviation to $M_i<M^{(1)}(0,0)$, the value of $F(M_i,0,0)$ will be greater than $1$, so the effective mass at the $i+1$th iteration of the SDE is given by
$M_{i+1}=M_iF(M_i,0,0)>M_i$. Therefore, if further iterative calculations are performed using the SDE, the effective mass will increase to $M_{i+1}<M_{i+2}<\cdots$. In this way, the effective mass obtained by the iterative method oscillates slightly around $M^{(1)}(0,0)$ and eventually converges to the solution $M^{(1)}(0,0)$. }

For any $T$ and $\mu$, if $F'(M,\mu,T)<0$ holds when $F(M,\mu,T)=1$, then a solution $M^{(1)}(T,\mu)$ should exist. If the initial input value is $M_0>M^{(1)}(T,\mu)$, it is expected that the effective mass will converge in the same way as when $T=\mu=0$ in the iterative calculation using the SDE.

The convergence properties of solutions to the SDE solved using iterative method are considered in more detail in Appendix B.

\vspace{5mm}

\vspace{5mm}

\begin{center}
{\large \bf Appendix B. Convergence of the solution using the SDE}
\end{center}

\vspace{2mm}

For any $T$ and $\mu$, let $M^{(1)}(T,\mu)$ be the solution corresponding to the effective mass at the local minimum of the effective potential, and assume that the initial input value $M_0$ satisfies $M_0>M^{(1)}(T,\mu)$.
We rewrite the SDE as
$$M_1=M_0F(M_0,\mu,T)=M_0-(1-F(M_0,\mu,T))M_0\equiv M_0-\Delta M_0 .$$
If $M_1>M^{(1)}(T,\mu)$,
$$ c(M_0,\mu,T) \equiv {\Delta M_0 \over M_0-M^{(1)}(T,\mu)}={(1-F(M_0,\mu,T))M_0 \over M_0-M^{(1)}(T,\mu)}<1$$
is satisfied.
When this condition holds for $M_1, M_2, \cdots$, then $M_i$ calculated iteratively by the SDE will converge from the region $M^{(1)}(T, \mu)<M_i$ to $M^{(1)}(T, \mu)$.

For any $M$, we write
$$ c(M,\mu,T) ={1-F(M,\mu,T) \over 1-M^{(1)}(T,\mu)/M}.$$

Since $M^{(1)}(T>0,\mu)<M^{(1)}(0,0)$ as shown in Appendix A, the denominator is $1-M^{(1)}(T>0,\mu)/M>1-M^{(1)}(0,0)/M$. Also, in $M/\Lambda\gg 1$, even for finite $T$ and $\mu$, the Boltzmann factor is $e^{\beta(E-\mu)}\gg 1$, so $\zeta(T,\mu,E(T,\mu,q))$ in Appendix A becomes $\zeta(T,\mu,E(T,\mu,q))\simeq 1$.
Therefore, for any $T$ and $\mu$, we obtain $F(M,\mu,T) \simeq F(M,0,0)$ for large $M$.
Using this approximation, we have
$$ c(M,\mu,T) < {1-F(M,\mu,T) \over 1-M^{(1)}(0,0)/M}\simeq {1-F(M,0.0) \over 1-M^{(1)}(0,0)/M}.$$
 When $m=M/\Lambda \gg 1$,
$$F(M,0,0)=2\pi\lambda\log{1+\sqrt{1+m^2} \over m}\simeq 2\pi\lambda/m.$$
Therefore, $c(M,\mu,T)<1$ is satisfied when $2\pi\lambda>m^{(1)}(0,0)$.
 For example, when $\lambda=0.3$, $2\pi\lambda\simeq 1.88$ and $m^{(1)}(0,0)\simeq 1.80$, so $2\pi\lambda>m^{(1)}(0,0)$ holds.

\begin{figure}
\centerline{\includegraphics[width=8cm]{CFMCONVM001B.ps}}
\caption{The effective mass dependences of  $c(M,\mu,T)$ for several different temperatures and chemical potentials  at $\lambda=0.3$. The horizontal axis shows the effective mass divided by $\Lambda$. }
%\label{Fig.22}
\end{figure}
Fig.22 shows the effective mass dependences of  $c(M,\mu,T)$ for several different temperatures and chemical potentials at $\lambda=0.3$. The accuracy of the calculation of $c(M,\mu,T)$ decreases when $M$ is close to $M^{(1)}(T,\mu)$. Therefore, in Fig. 22, the plot range of $M$ is limited to $M/\Lambda\geq 2>M^{(1)}(0,0)/\Lambda > M^{(1)}(T>0,\mu)/\Lambda$.
We will consider later the behavior of $c(M,\mu,T)$ as $M$ approaches $M^{(1)}(T,\mu)$.

In general, when $\lambda$ is small, $\xi=e^{1/(2\pi\lambda)}\gg 1$ holds. Also, since
$1/(2\pi\lambda)=\log\xi$, we obtain
$$ {m^{(1)}(0,0) \over 2\pi\lambda}={2\xi \over \xi^2-1}\log\xi \rightarrow 2{\log\xi \over \xi} \ll 1.$$

On the other hand, when $\lambda$ is large, we set $x=1/(2\pi\lambda)$ and
$$\xi=e^x=\sum_{n=0}^{\infty}{x^n \over n!}
=1+x\sum_{m=0}^{\infty}{x^m \over (m+1)!}\equiv 1+x\Sigma(x).$$
 From the above definition, we have 
$$ m^{(1)}(0,0)={2\xi \over \xi^2-1}={2(1+x\Sigma(x)) \over 2x\Sigma(x)(1+x/2\Sigma(x))}\equiv 2\pi\lambda K(x).$$
 Here, 
$$ K(x)={1+x\Sigma(x)\over \Sigma(x)(1+x/2\Sigma(x))}={1+x+x^2/2!+\cdots \over 1+x+11/12x^2+\cdots}$$
 gives $K(x)<1$ for $x\ll 1$.

\begin{figure}
\centerline{\includegraphics[width=8cm]{FMCONVM0GL001B.ps}}
\caption{The coupling constant dependence of the ratio $m^{(1)}(0,0)/(2\pi\lambda)$. The horizontal axis shows the coupling constant $\lambda$.}
%\label{Fig.23}
\end{figure}
Fig.23 shows the dependence of the ratio of the solution $m^{(1)}(0,0)$ and $2\pi\lambda$ on the coupling constant $\lambda$.
 As can be seen in Fig.23, the solution $m^{(1)}(0,0)$ satisfies $m^{(1)}(0,0)<2\pi\lambda$ for any $\lambda$, and its maximum value is $2\pi\lambda$. In addition, since the effective mass at finite temperature and finite density satisfies $m^{(1)}(T>0,\mu)<m^{(1)}(0,0)$, the maximum value of the effective mass corresponding to the local minimum of the effective potential is $2\pi\lambda$.
 
When $M$ is close to the solution $M^{(1)}(T,\mu)$, we get 
$$c(M,\mu,T)={(1-F(M,\mu,T))M \over M-M^{(1)}(T,\mu)}
\simeq -{(F(M,\mu,T)-F(M^{(1)},\mu,T))M^{(1)}(T,\mu) \over \delta M}$$
$$\simeq -{\delta F(M^{(1)},\mu,T)M^{(1)}(T,\mu) \over \delta M}
\rightarrow -F'(M^{(1)},\mu,T)M^{(1)}(T,\mu)=c(M^{(1)},\mu,T)$$
 by setting $M-M^{(1)}(T,\mu)\equiv \delta M$ and $F(M,\mu,T)-F(M^{(1)},\mu,T)\equiv \delta F(M^{(1)},\mu,T)$. 
If $c(M,\mu,T)$ is smaller than $1$, $M$ will converge to $M^{(1)}(T,\mu)$ in the region where $M^{(1)}(T,\mu)<M$.
From $-F'(M^{(1)},\mu,T>0)<-F'(M^{(1)},0,0)$ and $M^{(1)}(T>0,\mu)<M^{(1)}(0,0)$ as shown in Appendix A,
$$-F'(M^{(1)},\mu,T>0)M^{(1)}(T>0,\mu)<-F'(M^{(1)},0,0)M^{(1)}(0,0)$$ is satisfied. 
When $T=\mu=0$, we get 
$$F'(M^{(1)},0,0)=-2\pi\lambda\int_{0}^{\Lambda} dq{M^{(1)}(0.0) \over E^3(0,0,q)}$$
$$=-2\pi\lambda\int_{0}^{\Lambda} dq{M^{(1)}(0,0) \over (q^2+M^2(0,0))^{3/2}}
=-2\pi\lambda {\Lambda\over M^{(1)}(0,0)\sqrt{\Lambda^2+(M^{(1)}(0,0))^2}}.$$

From the above formula, we have
$$c(M^{(1)},0,0)= -F'(M^{(1)},0,0)M^{(1)}(0,0)=2\pi\lambda {\Lambda\over \sqrt{\Lambda^2+(M^{(1)}(0,0))^2}} =2\pi\lambda {1 \over \sqrt{1+(m^{(1)}(0,0))^2}}$$

For example, when $\lambda=0.3$, $m^{(1)}(0,0)\simeq 1.80$ gives
$$c(M^{(1)},0,0)=2\pi\lambda {1 \over \sqrt{1+(m^{(1)}(0,0))^2}}\simeq 0.92<1.$$

Fig.24 shows the chemical potential dependences of  $c(M^{(1)},\mu,T)$ at the solution $M^{(1)}(T,\mu)$ for several different temperatures at $\lambda=0.3$.

\begin{figure}
\centerline{\includegraphics[width=8cm]{FMCONVTMU001B.ps}}
\caption{The chemical potential dependences of  $c(M^{(1)},\mu,T)$ at the solution $M^{(1)}(T,\mu)$ for several different temperatures at $\lambda=0.3$. The horizontal axis shows the chemical potential divided by $\Lambda$.}
%\label{Fig.24}
\end{figure}

From Fig.24, it appears that $c(M^{(1)},\mu,T)<1$ holds for any $T$ and $\mu$ at $\lambda=0.3$.

In general, when $\lambda$ is small, $ m^{(1)}(0,0)$ satisfies $ m^{(1)}(0,0) \ll 1$.
This gives us $$c(M^{(1)},0,0)= {2\pi\lambda \over \sqrt{1+(m^{(1)}(0,0))^2}}\simeq 
2\pi\lambda \ll 1$$

On the other hand, $m^{(1)}(0,0)$ can also be written as $m^{(1)}(0,0) = 2\pi\lambda K(x)$, where $K(x)$ is given in Appendix A. Using this formula, we have
$$c(M^{(1)},0,0)= {2\pi\lambda \over \sqrt{1+(m^{(1)}(0,0))^2}}= 
{2\pi\lambda \over \sqrt{1+(2\pi\lambda K(x))^2}}.$$

Here, for large $\lambda$, $K(x)<1$ becomes very close to 1.
Therefore, for large but finite $\lambda$, we have $c(M^{(1)},0,0)<1$, and it is expected to asymptotically approach $1$ as $\lambda\rightarrow \infty$.

 Fig.25 shows The coupling constant dependence of $c(M^{(1)},0,0)$ at the solution $M^{(1)}(0,0)$. 

\begin{figure}
\centerline{\includegraphics[width=8cm]{FMCONVGL001B.ps}}
\caption{The coupling constant dependence of $c(M^{(1)},0,0)$ at the solution $M^{(1)}(0,0)$. The horizontal axis shows the coupling constant $\lambda$.}
%\label{Fig.25}
\end{figure}

 From the above considerations, for all $T$ and $\mu$, $M$ obtained by iterative calculation using SDE is expected to converge to $M^{(1)}(T,\mu)$ from the region where $M^{(1)}(T,\mu)<M$ in given $\lambda$.

Finally, we consider how to use iterative methods to check whether there are multiple solutions corresponding to the effective masses at local minima in the effective potential.
 If there are multiple solutions of $M^{(1)}(T,\mu)$ to the SDE, choosing $M^{(1)}(0,0)$ as the initial input will only converge to the largest solution of $M^{(1)}(T,\mu)$ that satisfies the SDE. When $F(M,\mu,T)>1$ for sufficiently small $M$, we use an iterative method to find a convergent solution starting from a small initial  value. If this solution coincides with the convergent solution $M^{(1)}(T,\mu)$ started from $M_0=M^{(1)}(0,0)$, then this is the only solution.
 If the two solutions do not match, then we need to iteratively find the convergence value by varying the initial input values for each $T$ and $\mu$ between the two solutions. On the other hand, when $F(M,\mu,T)<1$ for a sufficiently small $M$, if a small initial  value is input and an iterative method is used, the solution converges to $M(T,\mu) \rightarrow 0$. In this case, there exists at least one  spurious solution between the solution $M^{(1)}(T,\mu)$ obtained using $M_0=M^{(1)}(0,0)$ as the initial input value and $M=0$. Therefore, it is necessary to vary the initial input value from sufficiently small $M_0$ to  $M_0=M^{(1)}(T,\mu)$ obtained using $M_0=M^{(1)}(0,0)$ as the initial input value, then find the convergent solution at each $M_0$ for each $T$ and $\mu$.
 
 In order to know whether there are multiple convergent solutions, it seems to be efficient to calculate the $M$ dependence of $F(M,\mu,T)$ for each $T$ and $\mu$ in advance. As shown by the calculation results of $F(M,\mu,T)$ in the main text of the paper, our model has a unique solution $M^{(1)}(T,\mu)$ corresponding to the effective mass at a local minimum of the effective potential for each $T$ and $\mu$.

\vspace{2mm}

\vspace{5mm}
%-------------------------------------------------------------%                 %  REFERENCES
%%---------------------------------------------------------------------------
%\newpage
%\begin{thebibliography}{99}
\begin{flushleft} 
{\bf References }
\end{flushleft}

\vspace{2mm}

%\vspace{5mm}

\begin{description}
\item{[1]} R. Alkofer and L. von Smekal, The Infrared Behaviour of QCD Greens
Functions : Confinement, Dynamical Symmetry Breaking, and Hadrons as Relativistic Bound States, Phys.Rep.{\bf 353}, 281(2001) [arXiv:hep-ph/0007355 ].
\item{[2]} C. S. Fischer, Infrared properties of QCD from Dyson-Schwinger equations, J. Phys.G{\bf 32}, R253(2006)  [arXiv:hep-ph/0605173].
\item{[3]} M. Q. Huber, Nonperturbative properties of Yang-Mills theories, Phys.Rept.{\bf 879}, 1(2020) [arXiv:1808.05227 [hep-ph]].
\item{[4]} K. Nagata, Finite-density lattice QCD and sign problem: Current status and open problems, Prog. Part. Nucl. Phys. 127 (2022) 103991 [arXiv:2108.12423 [hep-lat]].
\item{[5]}  F. J. Dyson, The S Matrix in Quantum Electrodynamics, Phys.Rev.{\bf 75}, 1736(1949).
\item{[6]}  J. S. Schwinger, On the Green's Functions of Quantized Fields: I, Proc. Nat. Acad. Sci.{\bf 37}, 452(1951).
\item{[7]} D. J. Gross and A. Neveu, Dynamical symmetry breaking in asymptotically free field theories, Phys.Rev.{\bf D10}, 3235(1974).
\item{[8]} L. Jacob, Critical behavior in a class of O(N)-invariant field theories in two dimensions, Phys.Rev.{\bf D10}, 3956(1974). 
\item{[9]} B. J. Harrington and A. Yildiz, Restoration of dynamically broken symmetries at finite temperature, Phys.Rev.{\bf D11}, 779(1975).
\item{[10]} R. F. Dashen, S. Ma and R.Rajaraman, Finite-temperature behavior of a relativistic field theory with dynamical symmetry breaking, Phys.Rev.{\bf D11}, 1499(1975).
\item{[11]} S.-Z. Huang and M. Lissia, Contrasting real-time dynamics with screening phenomena at finite temperature, Phys.Rev.{\bf D53}, 7270(1996) [arXiv:hep-ph/9509360].
\item{[12]} B.-R. Zhou, Gap equations and effective potentials at finite temperature and chemical potential in D dimensional four-fermion models, Commun.Theor.Phys.{\bf 39}, 663(2003) [arXiv:hep-ph/0212193].
\item{[13]} J.-L. Kneur, M. B. Pinto, and R. O. Ramos, The 2d Gross-Neveu Model at Finite Temperature and Density with Finite N Corrections, Braz.J.Phys.{\bf 37}, 258(2007) [arXiv:0704.2843 [hep-ph]].
\item{[14]} M. Hayashi, T. Inagaki and W. Sakamoto, Phase Structure of a Four- and Eight-Fermion Interaction Model at Finite Temperature and Chemical Potential in Arbitrary Dimensions, Int.J.Mod.Phys.{\bf A25}, 4757(2010) [arXiv:1007.1497].\item{[15]} M. Le Bellac, Thermal Field Theory, Cambridge Monographs on mathematical physics, Cambridge university press (1996).
\item{[16]} M. Harada and A. Shibata, Chiral phase transition of QCD at finite temperature and density from the Schwinger-Dyson equation, Phys.Rev.{\bf D59}, 014010(1998) [arXiv:hep-ph/9807408].
\item{[17]} H. Tanaka and S. Sasagawa, Quark mass function at finite density in real-time formalism, Prog.Theor.Exp.Phys.{\bf 2020}, 053B06(2020) [arXiv:1907.02861 [hep-ph]].

\end{description}

\end{document}